\documentclass[12pt,a4paper]{article}
\usepackage{inputenc}
\usepackage{graphicx}
\usepackage{tabularx}
\usepackage{booktabs}
\usepackage{rotating}
\usepackage{multirow}
\usepackage{multicol}
\usepackage{amsmath}
\usepackage{amsthm}
\usepackage{placeins}
\usepackage{authblk}
\usepackage{natbib}

\usepackage{geometry}
\providecommand{\keywords}[1]
{
  \small	
  \textbf{\textit{Keywords---}} #1
}

\providecommand{\JEL}[1]
{
  \small	
  \textbf{\textit{JEL Codes:}} #1
}

\renewcommand{\vec}[1]{\mathbf{#1}}

\theoremstyle{definition}

\newtheorem{definition}{Definition}
\newtheorem{proposition}{Proposition}
\newtheorem{prediction}{Prediction}
\newtheorem{result}{Result}

\title{How Do Expectations Affect Learning About Fundamentals? Some Experimental Evidence \thanks{We thank Johannes Abeler, Bassel Tarbush, Stefania Innocenti, Miguel Ballester, Alex Teytelboym, Jens Koed Madsen, Giorgia Romagnoli, Theo Offerman and Sander de Vries, as well as seminar participants at the University of Oxford, for helpful comments. This project was approved by the Central University Research Ethics Committee at the University of Oxford. We obtained financial support from The Queen's College and Trinity College, University of Oxford. All remaining mistake are of course our own. Corresponding author: Kieran Marray. Email: kieranmarray@googlemail.com. Address: Tinbergen Institute Amsterdam, Gustav Mahlerplein 117, 1082 MS Amsterdam.}}
\author[1,2]{Kieran Marray}
\author[3]{Nikhil Krishna}
\author[4]{Jarel Tang}
\affil[1]{Institute for New Economic Thinking at the Oxford Martin School, University of Oxford}
\affil[2]{Tinbergen Institute}
\affil[3]{Trinity College, University of Oxford}
\affil[4]{The Queen's College, University of Oxford}
\date{}

\begin{document}

\maketitle

\begin{abstract}

We test how individuals with incorrect beliefs about their ability learn about an external parameter (`fundamental') when they cannot separately identify the effects of their ability, actions, and the parameter on their output.  \cite{Heidhues2018} argue that learning makes overconfident individuals worse off as their beliefs about the fundamental get less accurate, causing them to take worse actions. In our experiment, subjects take incorrectly-marked tests, and we measure how they learn about the marker's accuracy over time. Overconfident subjects put in less effort, and their beliefs about the marker's accuracy got worse, as they learnt. Beliefs about the proportion of correct answers marked as correct fell by $0.05$ over the experiment. We find no effect in underconfident subjects.
    
\end{abstract}


\JEL{D83, D84, D290}

\bigskip
\bigskip

An individual's output often depends not just on ability and actions, but other external factors or \textit{fundamentals}. To fix ideas, consider the example of a student completing an assignment. The number of answers they get correct depends on their ability and how much effort they put into studying. But their mark also depends on the proportion of these correct answers their professor marks as correct. Often, an individual cannot separately identify the effects of their ability and fundamentals on their output. An individuals' optimal action thus depends on their beliefs about their ability and the fundamental. Imagine the student only observes their mark (not individual questions, or the professor's feedback). They cannot separately identify the effect of the effort they put into studying or their natural ability on their mark from how harshly their professor marked the assignment. The student's beliefs determine their expected return to studying for the assignment, and thus what they think the optimal effort is.

 At the same time, many individuals' beliefs about their own ability are persistently incorrect (e.g see \cite{Svenson1981}, \cite{Camerer1999}, \cite{Hoffman2017}). Often individuals are overconfident (they overestimate their own ability at a given task), or undercondfident (they underestimate their own ability at a given task).  
 
 Here, we test how overconfidence and underconfidence affects learning about a fundamental when an individual cannot separately identify the effect of the fundamental and ability on their output. More specifically, we carry out an experimental test of \cite{Heidhues2018}, HKS18 hereafter. HKS18 characterise overconfident and underconfident individuals' equilibrium beliefs and learning process in these situations. They assume individuals only update their belief about the fundamental, and not their belief about their own ability, in response to output. Counter-intuitively, an overconfident individual's beliefs about the fundamental get less accurate, and their actions worse, the more they learn. The belief converges over time to a limiting belief that is very far from the true value. This happens because an overconfident individual always observes less output than they expect, attributes it all to the fundamental, and take a more sub-optimal action in response. The overconfident student blames lower than expected marks on the professor's marking, so concludes the returns to studying are lower than they actually are. As the student thinks the returns to studying are lower than they are, they put in too little effort into studying for the next assignment. Learning is self-defeating. Yet an underconfident individual's beliefs about the fundamental do not get so much worse as they learn. Instead, their beliefs about the fundamental converge to a region that is bounded by how incorrect their belief about their own ability is. This happens because an underconfident individual sometimes observes more output than they expect, but sometimes observes less, and still attributes it all to the fundamental. Learning process is self-correcting.
 
 Thus, the theory makes predictions about the overconfident and underconfident student's beliefs and actions. Overconfident students' beliefs about the proportion of correct answers their professor marks as correct should fall over time. Due to this, the effort they put into studying should also fall. The variance of underconfident students' beliefs should fall over time. Underconfident students move from an unrestricted prior beliefs to bounded posteriors. Beliefs are bounded below at the true proportion of correct answers the professor marks as correct. Thus, the effort they think is optimal will be higher than the overconfident student's, so their effort should fall by less than than the overconfident student's over time. The theory also gives a precise causal mechanism. Beliefs after one round cause changes in effort in the next round. Changes in effort lead to higher or lower output relative to expectations. Observing higher or lower output than expected then causes subjects to update their beliefs.
 
To test the theory, we experimentally induce the case of the student doing assignments. We get subjects to answer five sets of questions. These are marked by a `computerised marker' that marks a constant proportion of correct answers as correct, and we pay subjects a piece rate per mark. The fundamental is the proportion of correct answers marked as correct. Subjects only observe their mark each round. We record their effort by measuring how long it takes them to complete each set of questions. We elicit their beliefs about the proportion of correct answers the `computerised marker' marks as correct by getting them to bid for the chance to take an additional test after the experiment where each correct answer is marked as correct for sure. Subjects' optimal strategy is to bid their highest willingness to pay for the test -- the number of correct answers they expect to get given their current beliefs multiplied by the piece rate for correct answers. We can thus identify their beliefs about the proportion of correct answers the professor marked as correct by dividing the mark they observe by the number of correct answers implied by their bid.

Our design allows us to get clean estimates of the effect of misguided learning on beliefs about the fundamental. Under the theory, only the number of times subjects update their beliefs, their effort, and beliefs themselves change over the experiment. As we have an experimental setting, we can be sure that no other exogenous variables that might affect beliefs about the proportion of correct answers marked as correct change over the experiment.\footnote{A potential confound would be if subjects learnt how to better answer the questions over time. In the Appendix, we test for this, and find no evidence of it.} Thus, we can identify the causal effects of updating on these beliefs, and on actions. We can identify the direction of effects by comparing subjects beliefs about the proportion of correct answers marked as correct between the first and subsequent rounds. We can identify the magnitude of the effect by estimating a structural model for the effect of updating on beliefs about the proportion of correct answers marked as correct. We use the causal mechanism given in the theory to properly identify and control for the effects of subjects' prior beliefs each round.

 First, we test how subjects change their effort in response to feedback. The theory predicts that overconfident subjects will reduce their effort more over time than underconfident subjects. Next, we test for misguided learning in overconfident subjects. The theory predicts that updating should cause overconfident subjects' beliefs about the proportion of correct answers marked as correct to decrease. We compare distributions of beliefs after the first and fifth round with a one-sided paired t-test to find evidence of the sign. Then, to find the magnitude of misguided learning, we estimate a structural model for subjects beliefs about the proportion of correct answers marked as correct from the causal pathway given in the theory. The theory predicts a negative effect of updating on beliefs about the proportion of correct answers marked as correct conditional on prior beliefs. It also predicts that effort and past priors should be valid instruments for subjects' current priors.
 Finally, we repeat both of the methods above for underconfident subjects. Instead of looking at the level of their beliefs about the proportion of correct answers marked as correct, we look at their dispersion. The theory predicts that the dispersion should fall over time. In the appendix, we perform several robustness checks, and then use machine learning to look for indirect evidence for heterogeneous effects.

Our evidence is consistent with overconfident subjects learning about the fundamental is a misguided manner. The magnitude of the effect we find is small. We find no evidence that underconfident subjects learn like HKS18 predict. Overconfident subjects reduce the time spent on questions by $14.05$ seconds more each round than underconfident subjects. This is consistent with the causal mechanism HKS18 posit. When we compare the distributions of overconfident subjects' beliefs, we find no significant difference between beliefs after the first and final rounds. We do however find a significant negative difference between the beliefs after the first round and beliefs after all the other rounds. In our structural model, we find that updating causes overconfident subjects beliefs to fall by $0.01$. This corresponds to a fall of $0.05$ over the experiment. When we compare the dispersion of underconfident subjects' beliefs, we find no significant difference between beliefs after the first round and any other round. In our structural model, we find no effect of updating on the dispersion of beliefs.

Thus our results do suggest that overconfident individuals will act sub-optimally in these situations. But the effect size is relatively small. This is key because, if correct, 
HKS18's theory has important implications. Overconfidence is common and individuals cannot separately identify the effects of fundamentals from ability in a wide array of economically important situations, from CEOs making investment decisions (e.g see \cite{Malmendier2005} \cite{Malmendier2008}), to managers delegating to their employees (see \cite{Heidhues2018}), to partisan politicians choosing policies (e.g see \cite{Rollwage2018}). \footnote{See \cite{Heidhues2018} pp. $1165-1166$ for more applications} It suggests these individuals will often act sub--optimally because of how they learn, leading to large widespread welfare losses. Costly interventions to correct individuals expectations about their own ability could prevent this by changing how they learn, and thus be welfare-improving. Our results suggest the effect of such interventions will be small, so they might not be cost effective.
 
 Our study connects the emerging theoretical literature on mis-specified models with the large literature in on overconfidence. It is important to note that we differ from many papers in the literature in overconfidence in that we study an effect of overestimation of one's ability, as opposed to effects of overplacement of oneself relative to others (e.g see \cite{Burks2013} for effects in truck drivers, \cite{Park2010} for effects in poker players, \cite{Malmendier2005} for effects on CEOs), or excessive precision in one's beliefs (e.g see \cite{Soll2004}). These three are distinct, and not necessarily correlated \citep{Moore2008}.
 
 Our study is also one of the first empirical tests of how individual's learn with a mis-specified model. Thus, we build on the small theoretical literature in economics on behaviour with mis-specified models. \cite{Esponda2016} first formulate the equilibrium solution to a game with mis--specified models. \cite{Fundenberg2017} derive the limiting beliefs of individuals with incorrect priors. \cite{Hestermanforthcoming} set up a similar model to HKS18 but allow individuals' to endogenously choose their environment as well as their actions. They derive similar limiting beliefs for overconfident subjects, but do not find that the beliefs of underconfident subjects are bounded by their underestimation of their ability. Additionally, overconfident individuals select into high utility environments while underconfident individuals select into a low utility environments. This would be an interesting extension, but is outside the scope of our paper. \cite{Heidhues2020} apply a version of the HKS18 theory to explain endogenous emergence of discrimination. It also relates on self-image and ego-relevant parameters, as this is a plausible reason why individuals might maintain mis-specified models (e.g see \cite{Koszegi2006}).
 
 We are only aware of three other pieces of empirical work on learning with mis-specified models. \cite{Enke2020} studies how individuals form beliefs when information is selectively presented to them to corresponds to their priors. Subjects ignore selection, and form and maintain a mis-specified model of the environment. \cite{Coutts2020} show that in a team task, subjects distort information they receive to update more favourably than they should about themselves and a team-mate. A contemporaneous study by \cite{goette2021} tests how overplacement and underplacement affect learning about a fundamental in the HKS18 framework. They induce one example output function from the paper where HKS18 say their theory should apply (the case of a manager assigning tasks between themselves and a team), while we induce the other (the case of a student taking a test). They conduct a laboratory experiment, using a more abstract task (guessing the value of an integer drawn from $[-10, 10]$).  Like us, they find that overplacement causes misguided learning about the fundamental. But, unlike us, they find evidence for the theory amongst individuals who underplace themselves. As they test a different version of the theory (for overplacement as opposed to overestimation) in a different setting (the laboratory with students, as opposed to online with a more general population), our studies are complementary.
 
The rest of the paper proceeds as follows. 
Section $1$ describes our experiment design. Section $2$ describes HKS18's theory, and derives relevant predictions. Section $3$ presents our results. Section $4$ concludes, comparing our results in more detail to \cite{goette2021}. \\

\section{Design}

We aimed to create an environment where subjects' could not separately identify the effect of their ability and a fundamental on their output, and HKS18's theory should apply. Measuring how subjects' belief about the fundamental changes over the experiment then allows us to look for evidence of misguided learning. Subjects took five short verbal reasoning quizzes. A `computerised marker' marked a constant proportion of their correct answers as correct, and we paid subjects a piece rate per mark. After each test, subjects only see their mark. Thus, they have no way to separately identify the effect of their ability and the computerised marker on output. After they see their mark, we elicited subjects beliefs about the proportion of correct answers the marker marked as correct in an incentivised way. We pick a type of questions where the likelihood of getting a correct answer is known to be responsive to effort. Thus, if the theory is correct, subjects should exhibit misguided learning about the proportion of correct answers the computer marks as correct. This task is a key examples from HKS18 where they say misguided learning should occur (`output function 2'). We thus test the central prediction of the theory, using a prominent example in their paper.

 \begin{figure}[!h]
 \centering
   \caption{Experimental Setup}
\label{figure:1}
\includegraphics[width=15cm]{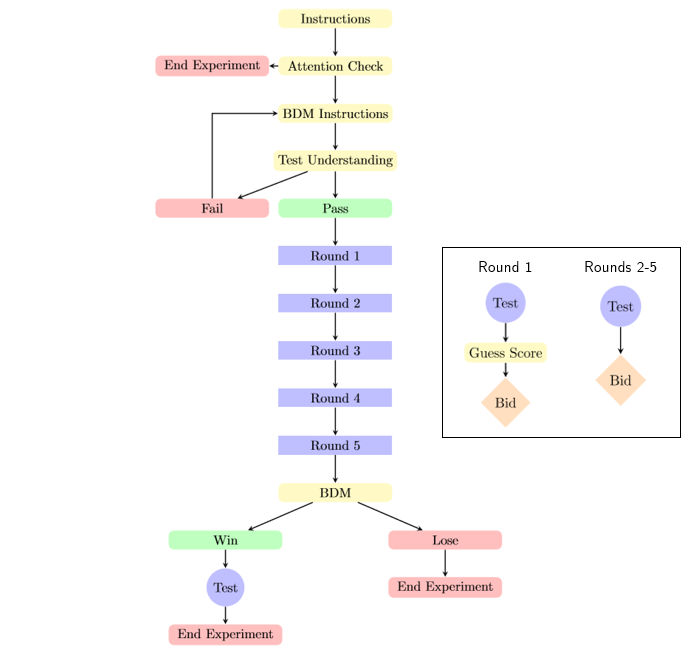}
 
 \end{figure}

Figure 1 gives an overview of the experiment.\footnote{We include all experimental materials in the Appendix}  We used a `within-sample' design - measuring each subject taking multiple tests over time as opposed to using a treatment and control sample (e.g see \cite{Charness2011}). This better allows us to see how subjects learn about the fundamental than using a treatment and control group. The experiment had two stages. Before the first stage, subjects read instructions. They were told that they would take a series of tests which `the computer' would mark, and would be paid a piece rate per mark. Furthermore, they were told that the computer would mark a fixed proportion of answers they got correct as incorrect, but not what that proportion was. This set priors over the fundamental as distributed over $[0,1]$. We embedded attention checks in the instructions so we could identify and exclude subjects who did not read them properly. This will reduce noise in our results

The first stage consists of five rounds. In each round subjects first answered eight questions. A computerised marker marked $50\%$ of their correct answers as correct, by dividing the number of answers they get correct by two. Subjects earn  $\$0.05$ per mark. After taking the test, subjects saw their marks. They next bid for the chance to take another test after the end of the experiment, with a piece rate of $\$0.20$ per mark. Crucially, in this final test, we told them that the computer will mark all correct answers as correct. Thus, they knew that their output in the final test would correspond to their true score.

In the first round, we also asked subjects
how many questions they had gotten correct before they saw their mark. If they are right, we pay them an additional $\$0.10$.  We used this to screen for overconfidence. If a subject thinks that they had gotten a higher number of questions correct than they actually had, we inferred that they were overconfident; if not, we inferred that they were underconfident.\footnote{Some might argue this is too weak a signal of overconfidence. But this is also used by others to determine whether individuals are overconfident (e.g \cite{Burks2013}). Some might also worry that ordering effects (e.g see \cite{Peiran2018}) will bias subjects' answers. But the distribution of subjects' answers and confidence provides strong evidence against ordering effects -- see the Appendix.} We only asked subjects once, after the first round of questions but before they saw their mark. Of course, subjects' confidence may have updated their belief about their ability over the course of the experiment This would be a reason that HKS18's theory may not apply, as in the theory misguided learning is driven by subjects not updating their belief about their ability. For now, we were only looking to test whether subjects exhibited misguided learning about the value of the fundamental. Thus, while changes in belief about ability might have occurred, they are outside the scope of our experiment. 

In the second stage, we randomly draw a bid for each subject from one round. We then use this bid to play a Becker-DeGroot-Marschak game (\cite{Becker1964}). The computer generates a random number between zero and $1.6$ (the maximum monetary value of an extra test to the subject). If a subject's bid is greater than or equal to the random number, they paid the value of the random number, took the extra test, and were paid $\$0.20$ for each answer they got correct. Else, the experiment ended. Subjects were aware that we would draw one of their bids at random to use in this auction game before they made any bids.

We observe the mark each subject gets each round. In the Becker-DeGroot-Maschak game a subject's best strategy was to bid the exact monetary value of the test to them when all correct answers are marked as correct  -- the score they think they will get given their belief about the marker that round.  Thus, we can recover the number of answers they think they would get correct given their beliefs as that time, by dividing their bid at $t$ by the value of each correct answer in the final test. Dividing the marks they got by the marks they think they would get if they took a test where all correct answers were marked as correct allows us to recover the fundamental. 

 We use a Becker-DeGroot-Maschak game instead of directly asking subjects their belief about the proportion of correct answers the computer marks as correct to ensure accurate measurement. Experimental tests have shown that, if it is stated carefully, subjects bid their maximum willingness to pay in BDM games not just in theory but in practice (see \cite{Lusk2007}). To ensure this will be the case, we took several steps. We based our instructions on the instructions in \cite{Berry2018}, who field tested and adjusted them to ensure subjects would understand the game correctly. We also used examples, and gave a link to a page explaining why it was in subjects' best interest to bid their maximum willingness to pay for the test. Finally, after explaining the mechanism and the optimal strategy, we quizzed subjects on how they should bid in a hypothetical scenario. Subjects could not progress to the first stage of the experiment until they answered correctly. If they answered incorrectly, they had to re-read the bidding instructions before answering again. These features have been shown to help subjects understand how to bid (see \cite{Lusk2007}). Hence each bid will be an accurate measure of subjects' willingness to pay. To further incentivise subjects to update on the information they have and bid accurately, we raised the piece rate for the final test to $\$0.20$ per question. 

We also took steps to prevent two potential issues with our questions: subjects looking up the answers online (and thus being able to separately identify the effect of the fundamental on output), or the test score being independent of effort. The first was a particular issue because we conduct the experiment on an online platform, where subjects commonly look up and then share answers to tasks on an associated forum. Thus, we took the questions from the practice book for the CEM $11+$ verbal reasoning examination: a verbal reasoning test taken by some primary school students in the UK. Subjects could not look up the answers to these particular questions online as they are only published in the practice book. Questions are not factual, and require subjects to think to obtain an answer. Experimental evidence shows that scores in these types of cognitive tests do vary with effort (e.g see \cite{Segal2012}, \cite{Borghans2008}). Thus, subjects could put in different levels of effort, and test scores would vary with effort. 

We recruited $226$ subjects on Amazon's `Mechanical Turk' platform. This is an online labour market, commonly used to recruit subjects for experiments, where individuals from all over the world complete tasks for a piece rate. The majority of participants on the platform come from India, and the United States. After excluding those who failed our attention check, $189$ remain.\footnote{Subjects who know that these studies often include attention checks may only pay attention to the bits of the experiment likely to include attention checks, and enter random results afterwards. To exclude such subjects, we remove those whose bids generate implausible values of $\phi$ - those that are undefined or greater than $1$} Our sample is balanced across age groups, ethnicities, and genders. The majority (145) were 21-35, though 29 were 36-50, 13 were greater than 50, and 2 were less than 21.  88 were female, 101 were male. 96 identified as `White, 77 identified as `Asian', 8 as `Black or African American', 2 as Native American, and 4 as `Other'. The experiment took an average of $37$ minutes including instructions. On average, subjects earned $\$0.81$, with standard deviation of $\$0.61$. We paid subjects an additional $\$1$ a participation fee. 

We picked our sample size and incentives to maximise our sample size and incentives we could provide given our budget and fairness concerns. We chose questions so that subjects would earn greater than the mean wage on Amazon Mechanical Turk in expectation ($\$3$ per hour at the time). This served two purposes. First, it was a way to ensure we paid subjects fairly given a small budget. In expectation, they would get at least their average outside option on the platform. Secondly, it ensured subjects have adequate incentives to perform well. Most Mechanical Turk tasks involve answering more questions, so subjects' payoffs per question are relatively large. To calibrate this, and ensure subjects got enough questions right to form a belief about the fundamental we conducted a pilot (n=$38$). We got subjects in the pilot to answer one of each of the type questions from our question set, and paid them for each answer they got correct. We then selected the types that the highest proportion of subjects got right, and used ones of this type in our main experiment.

\section{Predictions}

We now present the HKS18 model of how confidence leads to misguided learning, and draw out the testable predictions. First we describe the output function that corresponds to our task. This is a key example in the paper where their theory should apply. Then, we define our long-run equilibrium concept (following the exposition in \cite{Esponda2016}). This allows us to lay out the learning process and limiting beliefs for overconfident and underconfident subjects. We do not derive them mathematically as the proof is quite involved. We also restrict ourselves to the case where individuals myopically optimise and start with an accurate prior about the fundamental for ease of exposition. For more generality, see HKS18. Finally, we draw out the predictions theory makes. 

\subsection{Preliminaries}

 We can describe a subject using the output function:
\begin{equation*}
q_{t} = Q(e(\phi)_{t}(\phi), a, \phi) + \epsilon_{t},
\end{equation*}
where:\\
\begin{equation*}
Q(e_{t}, a, \phi) = \phi f(a,e(\phi)_{t}) - c(e).
\end{equation*}
\newline
$q_{t}$ corresponds to their utility. The number of answers a subject gets correct depends on their effort ($e$), and ability ($a$), which they combine in some production function $f(.)$.\footnote{We do not specify an exact functional form, as it not important for our analysis.} The computerised marker converts a proportion $\phi$ of the answers the subject gets correct into marks. $c(e)$ is a strictly convex effort cost, and $\epsilon_{t}$ is an i.i.d error term from a log--concave distribution.

We impose that $Q_{\phi)},Q_{e}, Q_{a}, Q_{e,\phi}>0$.  Output is increasing in the proportion of correct answers marked as correct, effort, and ability. The optimal amount of effort $e^{*}(\phi)$ is increasing in the proportion of correct answers the marker marks as correct.

Subjects make decisions in discrete time. Each period, they first choose to put in some amount of effort. They then observe only their output. They use this to update their beliefs about $\phi$ by Bayes rule. Subjects do not know the true $\phi$ and $a$, and start with heterogeneous priors over the full support $(0,1]$ . They only observe their output each time period. Thus, they cannot separately identify the effects of their effort, ability, and the computerised marker on output. 

Denote the individual's belief about their own ability as $\tilde{a}$. The key assumption is that subjects do not update $\tilde a$. Regardless of the marks they observe, subjects not change their belief about their ability, but only about how `harshly' the computerised marker marks them. This drives misguided learning.

\subsection{Equilibrium concept}

To characterise how subjects learn, we first need a way to characterise their steady state equilibrium beliefs. Our equilibrium concept formalises the notion that, in equilibrium, a subject's beliefs should be consistent with the output they observe given their belief about their ability. When they receive feedback, subjects update their belief about the proportion of correct answers the computerised marker marks as correct to obtain this consistency. HKS18 prove that the equilibrium is unique and that subjects converge to it as they learn.

To express this formally, we must first define a game with overconfidence. Our task is a simple game with a single player. A game with overconfidence comprises an objective game, and each subject's expected output function. This is a simplified version of the general game with mis-specified models in \cite{Esponda2016}. 

\begin{definition}
An objective game $\mathcal{O}$ is the tuple 
\begin{equation*}
\mathcal{O} = \langle I, e, A, \Phi, q, Q \rangle
\end{equation*}
\end{definition}

where I is the set of players, $e = \times_{i\in I}e_{i}$ is the set of actions, $a = \times_{i\in I}a_{i}$ is the set of actual abilities, $\phi$ are the fundamentals, $q = \times_{i\in I}q_{i}$ is the set of outputs, and $(Q^{i})_{i \in I}$ is a profile of output functions where $Q_{i}: e\times a \times \phi \rightarrow q^{i}$ maps actions, abilities, and fundamentals onto  outcomes. 

\begin{definition}
A set of expected output functions $\mathcal{Q}$ is the tuple 
\begin{equation*}
   \mathcal{Q} = \langle \tilde{a}, \tilde{\phi}, e, \tilde{Q}\rangle
\end{equation*}
\end{definition}
where $\tilde{a} = \times_{i\in I}\tilde{a}_{i}$ is the set of  beliefs about abilities, $\phi = \times_{i\in I}\phi_{i}$ is the set of beliefs about fundamentals, $e = \times_{i\in I}e_{i}$ is the set of actions, and $(\tilde{Q}^{i})_{i \in I}$ is a profile of expected output functions where $\tilde{Q}_{i}: e\times \tilde{a} \times \tilde{\phi} \rightarrow q^{i}$ maps actions, abilities, and fundamentals onto expected outcomes.

\begin{definition}
An subject's `surprise' is:\\
\begin{equation*}
\Gamma_{i}(\tilde{\phi}) = Q(e_{i}, a, \phi) - \tilde{Q}(e_{i}, \tilde{a}, \tilde{\phi}).
\end{equation*}
\end{definition}

 the difference between the output a subject gets given the true parameter values, and the output they expect to get given their beliefs about the parameter values, for a given action. Using this, we can define our solution concept.
\begin{definition}[Berk-Nash equilibrium]
A strategy profile $e^{*}$ is a Berk-Nash equilibrium of the game with overconfidence $\mathcal{G}=\langle \mathcal{O}, \mathcal{Q}\rangle$ if and only if for all $i\in I$, there exists an $\tilde{a}_{i}, \tilde{\phi}_{i}$ such that:
\begin{enumerate}
    \item $e^{*}_{i}$ is optimal given $\tilde{a}_{i}, \tilde{\phi}_{i}$, and
    \item $\Gamma_{i}(\tilde{\phi}) = 0$.
\end{enumerate}
\end{definition}

Another way of putting the second condition is that 

\begin{equation*}
Q(e_{i}^{*}, a_{i}, \phi)=\tilde{Q}(e_{i}^{*}, \tilde{a}_{i}, \tilde{\phi}_{i})
\end{equation*}

-- on average, the subject gets what they expect given their beliefs.\footnote{More generally, we should say that the Berk-Nash equilibrium the point that minimises the Kullback-Leibler divergence between the distribution of expected outcomes given their beliefs and the actual distribution of outcomes \cite{Esponda2016}. HKS18 show that this coincides with our simpler definition above for the game we consider.} Notice that a subject's beliefs do not have to be correct for this to be a stable equilibrium. They can be incorrect as long as expected output is consistent with actual output, so subjects have no reason to update their belief. Thus, the Berk-Nash equilibrium is stable.

\subsection{Overconfidence and misguided learning}

Now, we can characterise subjects' learning and equilibrium beliefs. First, consider subjects with $\tilde{a}>a$.

Overconfident subjects have a higher belief about their own ability than they should. They do not revise their beliefs about their own ability, so blame lower than expected output entirely on the fundamental. Optimal effort depends positively on beliefs about the fundamental, so when they revise their belief about the fundamental down it causes them to reduce their effort. This, and their overconfidence, means they always observe less output than they expect, causing them to revise their belief about the fundamental downwards and so on. This process continues until beliefs stabilise at the Berk-Nash equilibrium, where the subject gets the output they expect 

Our task is a simplified game with a single player. Thus, we drop the subscripts from above for ease of reading.  

HKS18 prove the following proposition.
\begin{proposition}[\cite{Heidhues2018}]
Consider an individual with $\tilde{a}>a$, who sets $e$ to myopically optimise $q$. Their belief $\tilde{\phi}_{i}$ converges to some $\phi_{\infty}$ such that $\phi_{\infty}<\Phi$. The action $e^{*}(\phi_{\infty})$ is a  unique, stable, Berk-Nash  equilibrium given beliefs $\tilde{a}$, $\phi_{\infty}$ .
\end{proposition} 

Now we sketch the `heuristic argument' from HKS18 that beliefs will converge to this Berk-Nash equilibrium goes as follows.\footnote{Note that our explanation looks quite different to the explanation in \cite{goette2021}. This is because they use `output function 1' from HKS18, which includes a loss function. We use `output function 2', which instead has effort costs.}. Figure 2.1 shows some possible time paths of beliefs from heterogeneous priors and levels of overconfidence.

Consider an overconfident subject in some time period t. As they are overconfident, $\tilde{a}>a$. For ease of exposition, assume that their initial belief about the proportion of correct answers the computerised marker marks as correct is accurate i.e $\tilde{\phi} = \phi$. Their expected output is 
\begin{equation*}
\tilde{Q}(e^{*}(\phi)_{t}, \tilde{a}, \tilde{\phi}) = \phi f(\tilde{a}, e^{*}(\phi)_{t}) - c(e^{*}(\phi)_{t}),
\end{equation*}

where if we take the first order conditions and set them equal to zero we can see that they set $e_{t}^{*}$ such that

\begin{equation*}
    \phi f'(\tilde{a}, e^{*}(\phi)_{t}) = c'(e^{*}(\phi)_{t}).
\end{equation*}

Their actual output is 

\begin{equation*}
Q(e^{*}(\phi)_{t}, A, \tilde{\phi}_{i}) = \phi f(a, e^{*}(\phi)_{t}) - c(e^{*}(\phi)_{t}).
\end{equation*}

We know that output is strictly increasing in ability, so the output they expect to get is higher than the output they actually get. $\Gamma(\phi) < 0$.

Now, when they observe this, the subject updates their beliefs. Remember that they do not update their belief about their own ability. Thus, they must update their belief about the fundamental downwards to some $\tilde{\phi}_{t+1}<\phi$ to account for the lower than expected output. The subject blames all of the lower-than-expected output on the computerised marker. 

In the next period, the subject therefore expects the following output
\begin{equation*}
\tilde{Q}(e(\tilde{\phi}_{t+1})_{t+1}^{*}, \tilde{a}, \tilde{\phi}_{t+1}) = \tilde{\phi}_{t+1}f(\tilde{a}, e(\tilde{\phi}_{t+1})_{t}^{*}) - c(e(\tilde{\phi}_{t+1})_{t}^{*}),
\end{equation*}

where they set $e(\tilde{\phi}_{t+1})_{t+1}^{*}$ such that

\begin{equation*}
    \tilde{\phi}_{t+1} f'(\tilde{a}, e^{*}(\tilde{\phi}_{t+1})_{t+1}) = c'(e^{*}(\tilde{\phi}_{t+1})_{t+1}).
\end{equation*}

We know that the optimal amount of effort is increasing in the belief about $\phi$. Thus,  revising their belief about $\phi$ down causes the subject to put in less effort than they should, given the true value of $\phi$. Intuitively, they think the returns to effort are lower than they actually are because they believe the computerised marker marks fewer of their correct answers as correct than it actually does.

Their actual output is 

\begin{equation*}
Q(e^{*}(\tilde{\phi}_{t+1})_{t+1}, a, \tilde{\phi}_{t+1}) = \phi f(a, e^{*}(\tilde{\phi}_{t+1})_{t+1}) - c(e^{*}(\tilde{\phi}_{t+1})_{t+1}).
\end{equation*}

Again, we know this must be lower than expected output from our conditions on the derivatives of the function. Thus, the subject observes less output than they expect. Again, they do not update their belief about their own ability. Instead, they blame the fundamental. Thus, in the next round they put in an even more sub--optimal level of effort, again get worse output than expected, and so on.

This will continue until subjects' beliefs and effort levels reach the Berk-Nash equilibrium. At this point, subjects observe the level of output they expect, so do not update their beliefs any further. Here, $\phi_\infty < \phi$, and $e^{*}(\phi_\infty) < e^{*}(\phi)$.

Subjects exhibit misguided learning  about $\phi$ in this model because they cannot update $\tilde{a}$ or separately identify the effects of $\phi$ from $a$ on output. Thus, the assumption subjects do not update $\tilde{a}$ is key - the degenerate belief is what drives their learning behaviour when they cannot separately identify effects. HKS18 do not just assume it because it generates a simple, tractable model. Crucially, they think it is an empirically realistic proxy for psychological forces that mean that individuals are often reluctant to revise beliefs about themselves. 

\begin{figure}
\caption{Example time path of beliefs about the fundamental of overconfident and underconfident agents who exhibit misguided learning}
  \begin{multicols}{2}
 \includegraphics[width=8cm]{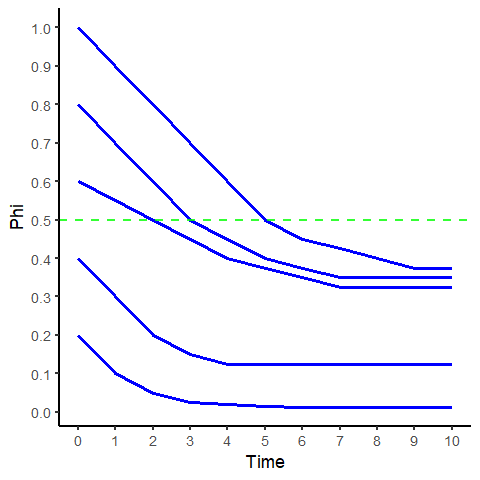}
 \includegraphics[width=8cm]{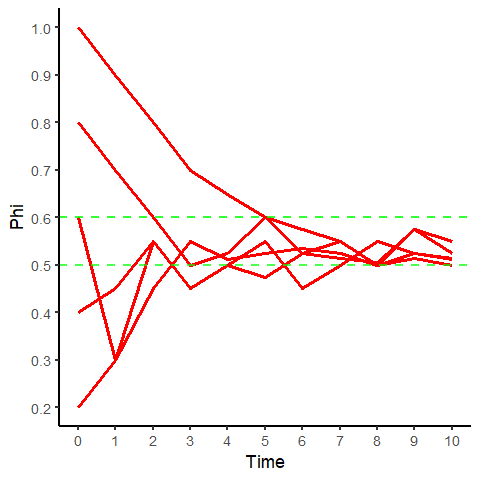}
 \end{multicols}
    \centering
  \footnotesize
Notes: Left panel gives example belief dynamics of overconfident agents from heterogeneous priors and beliefs in own ability. Right panel gives example belief dynamics of underconfident agents from heterogeneous priors where $a-\tilde{a} = 0.1$. The green line shows the true value of the fundamental (both panels) and upper bound on Berk-Nash equilibrium beliefs (right panel).
\end{figure}

\subsection{Underconfidence and bounded learning}

Now consider subjects with $\tilde{a}\leq a$. We group individuals whose beliefs about their own ability are accurate and those whose beliefs are about their own ability are too low together. For brevity, we refer to all of these individuals as `underconfident' (instead of using the term `underconfident and unbiased' as in \cite{goette2021}). Individuals with accurate beliefs are equivalent to underconfident agents, whose beliefs are bounded to end up being exactly accurate. 

Underconfident subjects have a lower belief about their own ability than they should. They do not revise their beliefs about their own ability, so attribute higher or lower than expected output entirely on the fundamental. Optimal effort depends positively on beliefs about the fundamental, so when they revise their belief about the fundamental down (up) it causes them to reduce (increase) their effort. Thus, they sometimes observe more output than they expect, and sometimes less than they expect. This adjustment process bounds their belief.

HKS18 prove the following proposition.
\begin{proposition}[\cite{Heidhues2018}]
Consider an individual with $\tilde{a}\leq a$, who sets $e$ to myopically optimise $q$. Their belief $\phi$ converges to some $\phi_{\infty}$ such that $|\phi_{\infty}-\Phi|\leq \Delta$, where $\Delta = a-\tilde{a}$. The action $e^{*}(\phi_{\infty})$ is a  unique, stable, Berk-Nash  equilibrium given beliefs $\tilde{a}$, $\phi_{\infty}$.
\end{proposition}

Again, we sketch the `heuristic argument'. Figure 2.2 represents some possible time paths of beliefs from heterogeneous priors.

Consider a strictly underconfident agent in period t. As they are strictly underconfident, their belief $\tilde{a}<a$. Their expected output is 
\begin{equation*}
\tilde{Q}(e^{*}(\tilde{\phi})_{t}, \tilde{a}, \tilde{\phi}) = \phi f(\tilde{a}, e^{*}(\tilde{\phi})_{t}) - c(e^{*}(\tilde{\phi})_{t}),
\end{equation*}

where if we take the first order conditions and set them equal to zero we can see that they set $e^{*}(\tilde{\phi})_{t}$ such that

\begin{equation*}
    \phi f'(\tilde{a}, e^{*}(\tilde{\phi})_{t}) = c'(e^{*}(\tilde{\phi})_{t})).
\end{equation*}

Their actual output is 

\begin{equation*}
Q(e^{*}(\tilde{\phi})_{t}), a, \phi) = \phi f(a, e^{*}(\tilde{\phi})_{t})) - c(e^{*}(\tilde{\phi})_{t})).
\end{equation*}

We know that output is strictly increasing in ability, so the output they expect to get is lower than the output they actually get. $\Gamma(\phi) > 0$.

Now, when they observe this, the subject updates their beliefs. Remember that they do not update their belief about their own ability. Thus, they must update their belief about the fundamental upwards to some $\tilde{\phi}_{t+1}>\phi$ to account for the higher than expected output. The subject attributes all of the higher-than-expected output to the computerised marker, not their own ability. 

In the next period, the subject therefore expects the following output
\begin{equation*}
\tilde{Q}(e^{*}(\tilde{\phi}_{t+1})_{t+1}, \tilde{a}, \tilde{\phi}_{t+1}) = \tilde{\phi}_{t+1} f(\tilde{a}, e^{*}(\tilde{\phi}_{t+1})_{t+1}) - c(e^{*}(\tilde{\phi}_{t+1})_{t+1}),
\end{equation*}

where they set $e^{*}(\tilde{\phi}_{t+1})_{t+1}$ such that

\begin{equation*}
    \tilde{\phi}_{t+1} f'(\tilde{a}, e^{*}(\tilde{\phi}_{t+1})_{t+1}) = c'(e^{*}(\tilde{\phi}_{t+1})_{t+1}).
\end{equation*}

We know that the optimal amount of effort is increasing in the belief about $\phi$. Thus,  revising their belief about $\phi$ upwards causes the subject to put in more effort than before. 

Now we can have two possible cases depending on how much greater : $\Gamma(\tilde{\phi}_{t+1})>0$, or $\Gamma(\tilde{\phi}_{t+1})<0$. In the first case, the underconfident subject responds by revising their belief about the computerised marker upwards and putting in more effort as above. In the second case, the underconfident subject responds by revising their belief about the computerised marker and their effort upwards as the overconfident subject always does. This second case can come about when the underconfident subject has revised their beliefs about the fundamental upwards by `too much  This process repeats until they reach a point where $\Gamma(\tilde{\phi})=0$ -- the Berk-Nash equilibrium. At this point, subjects observe the level of output they expect, so do not update their beliefs any further.

Here, $\phi_\infty > \phi$, and $e^{*}(\phi_\infty) > e^{*}(\phi)$. At this point, $Q(e_{i}^{*}, a_{i}, \Phi)=\tilde{Q}(e_{i}^{*}, \tilde{a}, \phi_\infty)$, so $|\tilde{\phi} - \Phi|$ depends on $|a - \tilde{a}|$. How underconfident the subject is bounds the distance of their final belief about the fundamental from its true value to a subset of the support between $\Phi$ and 1

\subsection{Testable predictions}

Start with overconfident subjects. We can see from the results above that the model makes the following predictions:

\begin{prediction}
If $\tilde{a}>a$
\begin{enumerate}
    \item $\tilde{\phi_{\tau}} < \tilde{\phi_{t}}$ for $\tau > t$ -- beliefs about the proportion of correct answers the computerised marker marks as correct fall over time;\
    \item $e^{*}_{\tau} < e^{*}_{t}$ for $\tau > t$ -- effort falls over time.
\end{enumerate}
\end{prediction}

We also have a clear causal pathway, which we can use to identify the size of the effect of misguided learning on beliefs. An overconfident subject puts in less effort than they should at t, because of their beliefs they form at t-1. This causes them to observe lower output than they expect at t. This then causes them to update their belief about the fundamental down at the end of t, which are based on their priors (belief at t-1) and the output they observe. Thus, we can use beliefs about $\phi$ at t-1, and effort at t to identify the size of the effect of updating at t. 

Now consider underconfident subjects. Again, we can see that the model makes the following predictions:

\begin{prediction}
If $\tilde{a}<a$
\begin{enumerate}
    \item $var(\tilde{\phi_{\tau}}) < var(\tilde{\phi_{t}})$ for $\tau > t$ -- the variance of subjects' beliefs about the proportion of correct answers the computerised marker marks as correct fall over time;\
    \item $var(e^{*}_{\tau}) < var(e^{*}_{t})$ for $\tau > t$ -- the variance of subjects' effort falls over time.
\end{enumerate}
\end{prediction}

The first follow because  subjects priors are distributed over the whole support $(0,1]$. As they update and converge towards equilibrium, their posteriors fall in a tighter interval until they can only have support in the interval $[\Phi, \Phi+\Delta]$. Thus, if we compute the variance across the whole population, it should fall over rounds. As optimal effort is a function of beliefs, its variance should also fall.

As a corollary from predictions 1 and 2, we get the following.

\begin{prediction} 
As they update, overconfident subjects will reduce their effort more than underconfident subjects.
\end{prediction} 

We should also see the following piece of indirect evidence.

\begin{prediction} 
 After multiple rounds of updating, overconfident individuals will have lower beliefs about $\phi$ on average than underconfident individuals.
\end{prediction}

The final follows from the others. Overconfident and underconfident subjects draw priors from the same distribution. Thus, before individuals start updating, $E(\phi_{Overconfident})=E(\phi_{Underconfident})$.  Overconfident individuals' beliefs fall as they update, while underconfident subjects' beliefs are bounded. Thus, the expectation of overconfident individuals' beliefs will be strictly lower than the expectation of underconfident individuals' beliefs. Note though that this is not decisive evidence for or against the theory, because confidence is not necessarily randomly assigned conditional on covariates.

We can use the structure of the output function to test these predictions. Take the output function, and separate the left hand side into observed output, the marks the student gets, and the cost of getting those marks. Subtracting $c(e)$ from both sides, we can define:

\begin{equation*}
\nu_{t} = \phi f(a,e_{t}).
\end{equation*}

 -- the output function in terms of the marks we observe $\nu$. We can thus recover their beliefs about the harshness of the marker by dividing the marks they got by the marks they think they should have gotten:

\begin{equation*}
\tilde{\phi}_{t} = \frac{\nu_{t}}{f(\tilde{a},e^{*}(\tilde{\phi}_{t}))}.
\end{equation*}

We can infer the mark they think they should have gotten from their bid in the Becker-DeGroot-Marschak game at t, adjusting for the higher piece rate in the final game.\footnote{A potential problem with this method is that, if $f(.)$ is concave and effort costs are convex, subjects will put in less than four times the effort in the final round compared to the first round (they cannot both be convex of course, as then no optimal effort level is guaranteed to exist). Thus, correcting for a difference in the price will lead to an overestimate of the $f(\tilde{a},e^{*}(\tilde{\phi}_{t})$, and thus an underestimate of $\tilde{\phi}_{t}$. This would increase any effects of misguided learning we observe. The same observation applies if individuals have a concave utility function over money and it is concave enough over the amounts of money the subjects earn in our experiment. Thus, our estimates can be seen as an upper bound on effects from misguided learning.} Thus, what matters for identification in this paper is that $\phi$ is multiplicative and $c(e)$ is separable. 

\section{Results}

\FloatBarrier

\subsection{Descriptive statistics}

Figures $3$ shows the distribution of overconfident and underconfident subjects' beliefs about $\phi$ (\textit{`Phi'}) each round. Table $1$ gives some descriptive statistics. Figure $4$ shows the mean of overconfident subjects' beliefs about $\phi$ each round. 

\FloatBarrier

 \begin{table}[!h]
   \caption{Subjects' Confidence, Raw Scores, Bids, and Beliefs About the Marker} 
 \centering
 \resizebox{1\width}{!}{\begin{tabular}{@{\extracolsep{5pt}}lccc} 
\\[-1.8ex]\hline 
\hline \\[-1.8ex] 
Statistic & \multicolumn{1}{c}{N} & \multicolumn{1}{c}{Mean} & \multicolumn{1}{c}{St. Dev.} \\ 
\hline \\[-1.8ex] 
Overconfidence & 189 & 2.450 & 2.422 \\ 
Score & 189 & 3.929 & 1.2752 \\  
Bid (implied score) & 189 & 5.205 & 0.780\\ 
Phi & 189 & 0.426 & 0.328\\  
\hline \\[-1.8ex] 
\end{tabular} 
}
 \begin{flushleft}
 \footnotesize
Notes: Score is the number of questions subjects got correct in a quiz (out of a possible 8 questions). Confidence is the score subjects thought they got in the first round before observing their mark minus the score they actually got. Bid is subjects' bids divided by $0.2$ (the dollar value of a correct mark). Phi is mark divided by bid.
\end{flushleft}
 \end{table}
 
 To see how beliefs about the fundamental vary by confidence and background characteristics, we regress subjects’ perceptions of the computer after round one and after round five on a dummy variable, $\text{Overconfident}$, which is $1$ if and only if $i$ is overconfident, and is $0$ otherwise, and gender, age, and ethnicity controls. 
  
  After round $1$:
\begin{equation*}
\phi_{i,1}= \alpha + \beta \cdot \text{Overconfident}_{i} + \gamma \cdot \vec{X} + u_{i}.
\end{equation*}
After round $5$:
\begin{equation*}
\phi_{i,5}= \alpha + \beta \cdot \text{Overconfident}_{i} + \gamma \cdot \vec{X} + u_{i}.
\end{equation*}

 \begin{figure}[!h]
    \caption{The Distribution of Overconfident (Blue) and Underconfident (Red) Subjects' Beliefs About the Accuracy of the Marker (`Phi') Each Round}
  \begin{multicols}{2}
 \includegraphics[width=8cm]{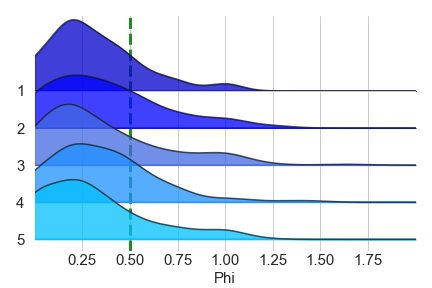}
 \includegraphics[width=8cm]{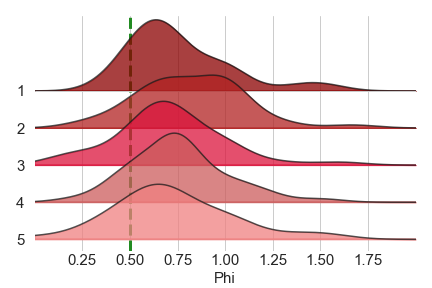}
 \end{multicols}
 \centering
  \footnotesize
Notes: The green line shows the true value of the fundamental. Distribution is calculated using a kernel-density estimate.
 \end{figure}
 
 \begin{figure}[!h]
\centering
   \caption{Mean Beliefs of Overconfident and Underconfident Subjects Over Rounds}
 \includegraphics[width=12cm]{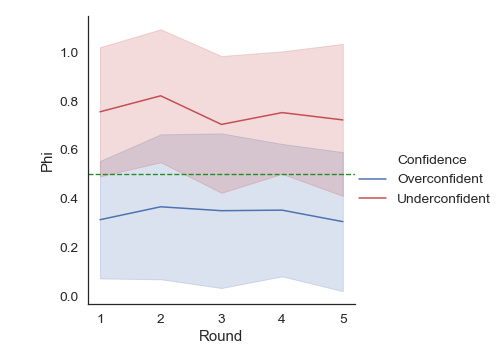}
 
 \footnotesize
 Notes: The green line shows the true value of the fundamental. Shaded areas show $2$ s.d around the mean.
 \end{figure}
 
 \FloatBarrier
 
The results of both regressions are given in Table $2$. There is a statistically significant difference between overconfident and underconfident subjects’ beliefs about $\phi$ after the final round. As the model predicts, overconfident subjects on average believe that the computer marked a lower proportion of their answers correctly than underconfident subjects. But we also find a statistically significant difference between overconfident and underconfident subjects' beliefs about the fundamental after the first round.  Figure $4$ shows this difference. The effect remains significant when we control for gender (dummy variable – 1 if male, 0 otherwise), age, and ethnicity (dummy variable – 1 if white, 0 otherwise), but gets smaller. All of the control variables are significant except age in the first round, and gender in both.

\begin{table}[!h]
\caption{The Difference Between Overconfident and Underconfident Subjects' Beliefs About the Accuracy of the Marker}
\centering
\resizebox{.75\width}{!}{\begin{tabular}{l c c c c c c c c}
\toprule
\toprule 
& \multicolumn{4}{c}{Dependent variable: $\phi_{i1}$} & \multicolumn{4}{c}{Dependent variable: $\phi_{i5}$}\\
& \multicolumn{4}{c}{(round $1$)} & \multicolumn{4}{c}{(round $5$)} \\
\cmidrule(l){2-9}&
\\
&(1)&(2)&(3)&(4)&(5)&(6)&(7)&(8)\\
\midrule
1 if overconfident&$-0.4422$***&$-0.4444$***&$-0.4318$***&$-0.4146$*** &$-0.4163$***&$-0.4172$***&$-0.3864$***&$-0.3577$***\\
&($0.044$)&($0.044$)&($0.044$)&($0.044$) &($0.052$)&($0.052$)&($0.049$)&($0.049$)\\[1ex]
1 if male& &$-0.0265$&$-0.0190$&$0.0052$& &$-0.0108$&$0.0074$&$0.0479$\\
& &($0.036$)&($0.036$)&($0.037$)& &($0.043$)&($0.040$)&($0.041$)\\[1ex]
Age& & &$0.0047$&$0.0031$& &  &$0.0114$***&$0.0088$***\\
& & &($0.002$)&($0.002$)& & &($0.002$)&($0.002$)\\[1ex]
$1$ if white & & & &$0.0902$**& & & &$0.1505$***\\
& & & &($0.039$)& & & &($0.044$)\\[1ex]
Constant &$0.7570$***&$0.7729$***&$0.6065$***&$0.5848$*** &$0.7232$***&$0.7297$***&$0.3233$***&$0.2871$***\\
&($0.039$)&($0.045$)&($0.084$)&($0.083$) &($0.046$)&($0.053$)&($0.094$)&($0.092$)\\[1ex]
Observations
&$189$&$189$&$189$&$189$&$189$&$189$&$189$&$189$\\
Adjusted $R^{2}$ &$0.349$&$0.348$&$0.363$&$0.377$&$0.253$&$0.249$&$0.337$&$0.374$\\
\bottomrule 
\addlinespace[1ex]
\end{tabular}}
\smallskip
\begin{flushleft}
\footnotesize
Notes: The dependent variable is the proportion of correct answers that the subject believes are marked as incorrect after the round given. 
Standard errors are given in parentheses.
Ex-ante power analysis suggested that with our sample sizes, we can detect the effect sizes we found with a power asymptotically close to $1$. 
\textsuperscript{***}Significant at the $1\%$ level\\ \textsuperscript{**}Significant at the $5\%$ level\\ \textsuperscript{*}Significant at the $10\%$ level\\
\end{flushleft}
 \end{table}

We do not try to identify a causal relationship - there might ommitted variable bias from factors we did not observe, such as IQ (e.g see Burks et al. $2013$). 

These results are consistent with prediction 4.  If the theory is correct, overconfident subjects' beliefs about $\phi$ should fall over the rounds. By contrast, underconfident subjects' beliefs about $\phi$ should converge towards $0.5$ over the rounds. Thus, we should find a significant difference between overconfident and underconfident subjects' beliefs in the final round.

\subsection{Effort provision}

Our first result provides support for prediction $3$.

\begin{result}
Overconfident subjects reduce their effort by more over time than underconfident subjects.
\end{result}

We measure effort by looking at how long subjects spend on each round of questions. As we use a within--subjects experiment, we cannot test for reduction in effort by just comparing means over time. We would expect the time subjects' spend to fall over the experiment even without misguided learning because they become more familiar with the experimental setup. Instead, we look at the difference in the effect of updating on overconfident and underconfident subjects. If subjects do not learn in a misguided manner, there should be no difference in the effect of updating on overconfident and underconfident subjects' effort.

To test this, we run the random--effects panel regressions of the effect of updating on effort provision. As, by construction, we have no individual fixed effects that are correlated with the round number, our estimated coefficient will be causal estimates.

\begin{table}[!htbp] \centering 
  \caption{The effect of updating on effort provision} 
  \label{} 
\begin{tabular}{@{\extracolsep{5pt}}lccc} 
\\[-1.8ex]\hline 
\hline \\[-1.8ex] 
 & \multicolumn{3}{c}{\textit{Dependent variable: $\text{Effort}_{it}$}} \\ 
\cline{2-4} 
\\[-1.8ex] & (1) & (2) & (3)\\ 
\hline \\[-1.8ex] 
 $\text{Round}_{it}$ & $-$47.11$^{***}$ & $-$33.06$^{***}$ & $-$33.06$^{***}$ \\ 
  & (4.66) & (4.83) & (8.29) \\ 
  & & & \\ 
 Overconfident? &  &  & 42.12 \\ 
  &  &  & (35.03) \\ 
  & & & \\ 
 $\text{Round}_{it} \times \text{Overconfident?}$  &  &  & $-$14.05$^{*}$ \\ 
  &  &  & (9.34) \\ 
  & & & \\ 
 Constant &  &  & 345.70$^{***}$ \\ 
  &  &  & (31.11) \\ 
  & & & \\ 
\hline \\[-1.8ex] 
N & 745 & 200 & 945 \\ 
\hline 
\hline \\[-1.8ex] 
\textit{Note:}  & \multicolumn{3}{r}{$^{*}$p$<$0.05; $^{**}$p$<$0.01; $^{***}$p$<$0.001} \\ 
 & \multicolumn{3}{r}{} \\ 
\end{tabular} 
\smallskip
\begin{flushleft}
\footnotesize
Notes: The dependent variable is the amount of time the subject takes to complete the questions in the round given.
Standard errors are given in parentheses. All values are rounded to two decimal places.\\
\textsuperscript{***}Significant at the $1\%$ level\\ \textsuperscript{**}Significant at the $5\%$ level\\ \textsuperscript{*}Significant at the $10\%$ level\\
\end{flushleft}
\end{table}

We report the results in Table 3. We estimate the first two models on overconfident and underconfident subjects individually. In the third regression, we pool our sample. When we interact a dummy overconfidence with number of updates, we find a negative coefficient that is statistically significant at the $10\%$ level. Overconfident subjects reduced their effort at a greater rate than underconfident subjects over the experiment. 

\subsection{Overconfident subjects}

We find weaker evidence for prediction 1.

\begin{result}
Updating has a small negative effect on overconfident subjects' beliefs about the proportion of correct answers marked as correct.
\end{result}

We first test to see if there is a difference between overconfident subjects' beliefs about the proportion of correct answers marked as correct over rounds. If the theory is correct, we should find a significant negative difference between beliefs in subsequent rounds. As in \cite{goette2021}, we conduct pairwise one--sided t-tests to test for this. We find no evidence of a difference between beliefs in the first and fifth round, but it is not statistically significant ($0.008$, p-value: $0.654$). We do find negative differences between beliefs in the first and fourth round ($-0.039$, p-value: $0.013$), first and third round ($-0.037$, p-value: $0.032$), and first and second round ($-0.037$, p-value: $0.031$). The size of these differences are relatively small. 

These results are still consistent with a small but significant effect of misguided learning. Subjects might learn in a misguided manner, but the effect size may not be big enough that we detect large differences in means. 

We can use a structural panel model to look for such small effects. As we have experimental data, we know that the only thing that changes over the rounds is the number of times the subject updates their belief, and their priors.\footnote{Another thing that could change is that subjects could learn how to answer the questions, and hence get better at the test. This would invalidate our identification strategy. We carry out a robustness check where we test for learning effects (see Appendix), and find no evidence that subjects got better at tests over the course of the experiment.} Thus, by estimating a panel model conditioning on individuals' priors each round (an Arrelano-Bond estimator), we can estimate the causal effect of updating on beliefs. The model gives us a specific causal pathway from effort to beliefs. Thus, we can use effort levels and past priors as our instruments for the current prior to identify our coefficients. If the theory is correct, these models should be correctly specified (i.e we find evidence for the causal pathway in the model) and we should find a significant negative effect of updating on beliefs.

Our structural model is of the form

\begin{equation*}
\phi_{i,t} = \alpha + \beta \cdot Round_{i,t} + \gamma \phi_{i,t-1} + u_{i,t}
\end{equation*}

 Including individual fixed effects controls for all time-invariant individual differences. The coefficient $\beta$ is the effect of just changing the number of times they have updated on subjects' beliefs about $\phi$, conditional on the previous period's belief. As we condition on the previous period's belief, this isolates the pure effect of updating more times on beliefs. As, by construction, we have no individual fixed effects that are correlated with the round number, our estimated coefficient will be causal estimates of the effect of updating on beliefs.

\begin{table}[!htbp] \centering 
  \caption{Structural models of the effect of updating on overconfident subjects' beliefs.} 
  \label{} 
\begin{tabular}{@{\extracolsep{5pt}}lccc} 
\\[-1.8ex]\hline 
\hline \\[-1.8ex] 
 & \multicolumn{3}{c}{\textit{Dependent variable: $\Delta \phi_{it}$}} \\ 
\cline{2-4} 
\\[-1.8ex] & (1) & (2) & (3)\\ 
\hline \\[-1.8ex] 
 $\Delta \text{Round}_{it}$ & $-$0.02$^{***}$ & $-$0.01$^{**}$ & $-$0.02$^{***}$ \\ 
  & (0.01) & (0.01) & (0.01) \\ 
  & & & \\ 
 $\Delta \phi_{it-1}$ & 0.34$^{**}$ & $-$0.33$^{***}$ & 0.33$^{**}$ \\ 
  & (0.14) & (0.05) & (0.14) \\ 
  & & & \\ 
\hline \\[-1.8ex] 
Instruments: & & &\\
$\Delta$ $\text{Effort}_{it-1}$:& No & Yes & Yes\\
$\Delta$ $\phi_{it-2}$: & Yes & No & Yes \\
\hline 
Diagnostic tests: & & &\\
J statistic: & 0.11 & 5.78 & 7.08\\
First-stage F statistic: & 14.97 & 66.73 & 14.44\\
Autocorrleation test (1): & -3.56 & -2.22 & -3.60\\
Autocorrleation test (2): & -1.34 & -2.75 & -1.34\\
\hline
N & 447 & 447 & 447 \\ 
\hline 
\hline \\[-1.8ex] 
\end{tabular} 
\smallskip
\begin{flushleft}
\footnotesize
Notes: The dependent variable is the proportion of correct answers that the subject believes are marked as incorrect after the round given. Effort is the amount of time the subject takes to complete the questions in the round given. For identification, we estimate the model in first-differences.
Standard errors are given in parentheses. All values are rounded to two decimal places.\\
\textsuperscript{***}Significant at the $1\%$ level\\ \textsuperscript{**}Significant at the $5\%$ level\\ \textsuperscript{*}Significant at the $10\%$ level\\
\end{flushleft}
\end{table}

We report the results of an Arellano-Bond estimator of this structural model in Table 2. We report three specifications. In the first, we only use past effort as an instrument for subjects' current beliefs. In the second, we only use past beliefs as an instrument for subjects' current beliefs. In the third, we use both past effort and past beliefs as instruments for current beliefs. We use these because these correspond to the causal mechanism in the model. We report the relevant specification checks in the bottom rows of the table. All models fail to reject the null hypothesis in the Sargan test, have with high first--stage F statistics, and are only correlated at the correct lags. Thus, we have evidence to suggest that all of these models are correctly specified. The first--stage F-statistic is highest in the model where we only use effort as an instrument. Thus, this is our preferred specification.

In each model, we find a small negative coefficient on number of updates, which is highly statistically significant. This suggests a small negative causal effect of updating on beliefs about the proportion of correct answers marked as correct. We find a much larger coefficient on individuals priors in each model, also statistically significant. Looking at the size of the coefficient, just updating causes overconfident subjects' beliefs about the proportion of correct answers marked as correct fall by $0.05$ of the course of the experiment.

\subsection{Underconfident subjects}

We find do not find any decisive evidence for prediction 2.

\begin{result}
Updating has no effect on the spread of underconfident subjects' beliefs about the proportion of correct answers marked as correct.
\end{result}

We first test to see if there is a difference between the squared distance of underconfident subjects' beliefs about the proportion of correct answers marked as correct over rounds. We use squared distance because the model predicts that underconfident subjects' beliefs will converge to some distance $\Delta$ from $\phi$. They may approach this point from below, above, and we want to treat these equally. If the theory is correct, we should find a significant negative difference between beliefs in subsequent rounds (corresponding to a difference between the variance of beliefs, as the mean of the squared differences is the variance).

Following the analysis in \cite{goette2021}, we conduct pairwise one--sided t-tests. We find a negative difference between the first and fifth round, but it is not statistically significant ($-0.026$, p-value: $0.146$). We find no significant differences between the first and earlier rounds either. These results are robust to using absolute differences from mean beliefs instead of squared differences ($-0.031$, p-value: $0.171$). We do find a significant effect when we restrict ourselves to strictly underconfident subjects ($-0.039$, p-value: $0.042$).

Again, we use a structural panel model to identify the size of any effect as for overconfident subjects.

\begin{equation*}
y_{i,t} = \alpha + \beta \cdot \text{Round}_{i,t} + \gamma y_{i,t-1} + u_{i,t},
\end{equation*}

where $y_{it} = (\phi_{it} - \frac{\sum_{i=1}^{N}\phi_{it}}{N})^{2}$.

We report the result in Table 3.

\begin{table}[!htbp] \centering 
  \caption{Structural models of the effect of updating on underconfident subjects' beliefs.} 
  \label{} 
\begin{tabular}{@{\extracolsep{5pt}}lccc} 
\\[-1.8ex]\hline 
\hline \\[-1.8ex] 
 & \multicolumn{3}{c}{\textit{Dependent variable: $\Delta y_{it}$}} \\ 
\cline{2-4} 
\\[-1.8ex] & (1) & (2) & (3)\\ 
\hline \\[-1.8ex] 
 $\Delta\text{Round}_{it}$& $-$0.004 & $-$0.01 & $-$0.002 \\ 
  & (0.01) & (0.01) & (0.01) \\ 
  & & & \\ 
 $\Delta y_{it-1}$& $-$0.05 & $-$0.47$^{***}$ & $-$0.05 \\ 
  & (0.16) & (0.18) & (0.16) \\ 
  & & & \\ 
\hline \\[-1.8ex] 
Instruments: & & &\\
$\Delta$ $\text{Effort}_{it-1}$:& No & Yes & Yes\\
$\Delta$ $y_{it-2}$: & Yes & No & Yes \\
\hline 
Diagnostic tests: & & &\\
J statistic: & 10.49 & 6.79 & 6.72\\
First-stage F statistic: & 0.19 & 0.36 & 8.86\\
Autocorrleation test (1): & -1.83 & -1.83 & -1.56\\
Autocorrleation test (2): & 1.75 & 1.75 & 0.54\\
\hline
N & 120 & 120 & 120 \\ 
\hline 
\hline \\[-1.8ex] 
\end{tabular}
\smallskip
\begin{flushleft}
 \footnotesize
Notes: The dependent variable is the squared difference between the proportion of correct answers that the subject believes are marked as incorrect after the round given and the mean across indviduals. Effort is the amount of time the subject takes to complete the questions in the round given. For identification, we estimate the model in first-differences.
Standard errors are given in parentheses. All values are rounded to two decimal places.\\
\textsuperscript{***}Significant at the $1\%$ level\\ \textsuperscript{**}Significant at the $5\%$ level\\ \textsuperscript{*}Significant at the $10\%$ level\\
\end{flushleft}
\end{table}

We report the result in Table 3. In each model, we find a small negative coefficient on number of updates, but it is not statistically significant. Again, we find the same using absolute deviations instead of squared deviations. Notice that each of these structural models fails almost all of the standard robustness checks for instrumental variables. All the models reject the null hypothesis in Sargan test, have first-stage F statistics below $10$, and fail the autocorrelation tests for validity of the Arrelano-Bond identification strategy. This suggests that the causal structure in HKS18 is not correct for underconfident subjects.

\FloatBarrier
 
\section{Discussion}

In this paper, we test the theory in HKS18 of that confidence affects how individuals learn about fundamentals when they cannot separately identify the effects of their ability and the fundamentals on their output. This describes a large proportion of individuals in many economically important situations, such as managers delegating to employees, politicians allocating resources to projects, and CEOs making investment decisions. If their theory is correct, overconfident individuals act sub--optimally in all of these situations because of how they learn about the fundamentals. As these decisions affect many individuals and a huge percentage of world output, this would imply large welfare losses. If a social planner could intervene to correct how they learn for less than the costs the decisions impose, perhaps by review processes that force individuals to update their beliefs about their own ability or allow them to separately identify its effect on output, the theory implies that this would be welfare--improving.

Our results are consistent with the contemporaneous study by \cite{goette2021}. We test a different output function from the theory in a different setting with a different population using similar, but different, methods. Thus, our results are complementary. \cite{goette2021} experimentally test the effect of overplacement and underplacement on learning about a fundamental in the HKS18 framework. We test the effect of overestimation and underestimation instead. They use a more abstract task where subjects guess the values of integers drawn from $[-10,10]$, in a laboratory experiment with undergraduate students. We use a more concrete task, in a setting closer to the field, and with a population with a wider range of ages and backgrounds. To test for an effect of updating on beliefs about the fundamental, they also conduct pairwise t-tests on distributions of beliefs across rounds. They find a significant difference in beliefs between some but not all of rounds, as we do. They do not structurally estimate the size of the effect of misguided learning as we do, but instead look at the of learning on effect of bias by comparing subjects' beliefs to the true state in each round. They find a small but significant effect of misguided learning for both overconfident and underconfident subject. Another difference is that they use higher absolute incentives. Given that we calibrate our incentives to be large relative to our subjects' outside options, this should not lead to any significant difference in results.

As a robustness exercise, we replicate our analysis for underconfident subjects using the difference between their beliefs and the true state as in \cite{goette2021}. This allows us to see if our lack of significant results are driven by our choice of measure. We do now find significant differences between the dispersion of underconfident beliefs after the first round and after all subsequent rounds apart from the final rounds (results in order from final to second round: $-0.049$, p-value:$0.181$; $-0.057$, p-value:$0.036$; $-0.073$, p-value:$0.049$; $-0.087$, p-value: $0.006$). Our results from the structural estimation do not change.

Overconfident subjects learn in a way that is consistent with misguided learning. Underconfident subjects did not learn in a way that is consistent with misguided learning. In overconfident subjects, the effect sizes of misguided learning are small, however. An explanation of this could be that subjects partially update their beliefs about their own ability.  Experimental evidence that overconfident individuals do seek information about their own ability when updating their beliefs supports this explanation (e.g see \cite{Burks2013}). For example, a reason individuals might not update their beliefs about their own ability in some situations because they get utility from positive self image (e.g \cite{Koszegi2006}). Subjects may have gotten little ego utility from their belief about their ability to do the tests we set them. If that was the case, subjects would be relatively willing to update their beliefs about their ability, leading to only a little misguided learning. This is also consistent with the results of the final treatment in \cite{goette2021}, where they find larger effects when the measure of ability is more ego-relevant. The model could also miss important causal channels that affect learning about the fundamental. Though misguided learning seems to be a channel, factors that covary with oveconfidence could directly cause individuals to have unrealistically low perceptions of the fundamental. This explanation is also consistent with experiments that have shown that sub-optimal decisions made by overconfident individuals are sensitive to factors other than confidence (e.g \cite{Malmendier2005}).

Others could extend our research by directly testing this explanation. An intresting extension would be to see whether misguided learning occurs in the field. If lack of updating of beliefs about a parameter could be driven by ego utility, we would expect this to be stronger in situations where an individual's reputation is on the line (e.g amongst CEOs or politicans) than in an experiment. This would also allow one to compute the magnitude of welfare losses from misguided learning. 
\newpage


\bibliographystyle{agsm}
\bibliography{Unrealbib}

\newpage
 
\section{Appendix}

\subsection{Robustness Checks - Order Effects}

Ordering effects (e.g see \cite{Peiran2018}) might confound our measure of subjects' overconfidence, as subjects answers may depend on our framing in the instructions. To assess whether subjects are overconfident or underconfident, we ask to guess the number of questions they actually got correct in the first round before we tell them their mark, and compare that to the number of questions they did actually get correct. In our instructions, we say that the marker will mark some correct answers as incorrect. This is not directly relevant to subjects' answers, as we clearly ask them about the questions they actually got correct, not the mark that the marker gives them. This might cause an order effect, however -- our negative framing of the marker in the instructions may spill over into subjects perceptions of their own ability. If so, subjects estimates of their score will be lower than they should be. Thus, we would infer that more subjects are underconfident than we should. As we are seeking to identify different effects in both groups, this would distort our results.

\FloatBarrier

\begin{figure}[!h]
 \begin{multicols}{2}
   \caption{Subjects' Predicted Scores}
 \includegraphics[width=8cm]{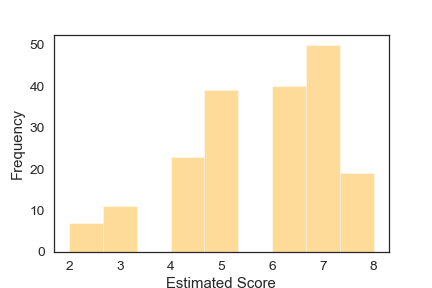}
 
   \caption{Subjects' Confidence}
 \includegraphics[width=8cm]{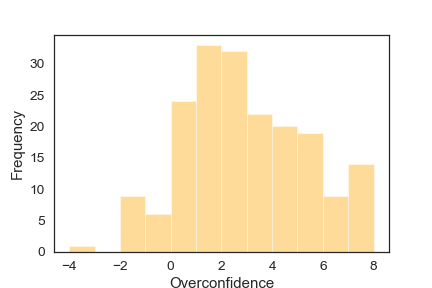}
 \end{multicols}
 \end{figure} 

\FloatBarrier

Figure $7$ shows the distribution of subjects beliefs about the number of questions they got correct. If there are these order effect in our experiment, our framing of the instructions will shift individuals' beliefs about their own ability downwards from what they should have been. Thus, we would not expect to see many individuals with very high beliefs about their own ability. But, many subjects do have high beliefs about their own ability - $23$ out of $189$ think they got all of the questions correct, and $79$ think they only got one question wrong. This is strong evidence we do not have order effects.

\subsection{Robustness Checks - Learning Effects}

As we use a within-subject design, we carry out some further analysis to see if there is evidence that subjects learn how to answer our tests over the course of the experiment (\textit{learning effects}). Learning effects are a common problem for within-sample designs, and would cast doubt on our results. We construct the following fixed effect regression model for overconfident and underconfident subjects:\footnote{We carry out a Hausman test to determine whether a random or fixed effects model was appropriate. We are able to reject the null hypothesis that a random effects model is appropriate at a $5$ percent significance level $(p=1.98.e^{47})$.}

\begin{equation*}
    SCORE_{i,t}= \alpha_{i} + \beta.ROUND_{i,t} + u_{i,t}
\end{equation*}

If subjects learn the answers to questions over the course of the experiment, we would expect to see a significant positive coefficient on scores in each regression.\\

The results of the regressions are presented in Table $3$. The mean scores of overconfident and underconfident subjects each round are shown in Figures $9$ and $10$.\\

\FloatBarrier
\begin{table}[!h]
\caption{The Effect of Taking Tests on Subjects' Actual Test Scores}
\centering
\resizebox{1\width}{!}{\begin{tabular}{l c c}
\toprule
\toprule 
& Fixed Effects: & Fixed Effects\\
&Test Score- & Test Score-\\ 
&Overconfident Subjects & Underconfident Subjects\\
\midrule
&(1)&(2)\\
\midrule
Round& $0.0322$ & $-0.1675$*** \\
& ($0.0293$)&($0.0430$)\\[1ex]
Observations&$745$&$200$\\
$R^{2}$ &$0.0023$&$0.0617$\\
\bottomrule 
\addlinespace[1ex]
\end{tabular}}
\begin{flushleft}
\smallskip

\footnotesize
Notes: The dependent variable is subjects' actual score in the test each round. Standard errors are given in parentheses.

\textsuperscript{***}Significant at the $1\%$ level\\
\textsuperscript{**}Significant at the $5\%$ level\\ 
\textsuperscript{*}Significant at the $10\%$ level\\
\end{flushleft}
 \end{table}
 
\begin{figure}[!h]
 \begin{multicols}{2}
   \caption{Overconfident Subjects' Scores Over the Experiment}
 \includegraphics[width=6cm]{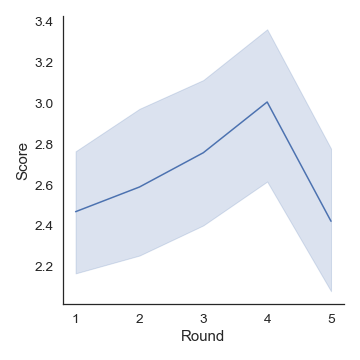}
 
   \caption{Underconfident Subjects' Scores Over the Experiment}
 \includegraphics[width=6cm]{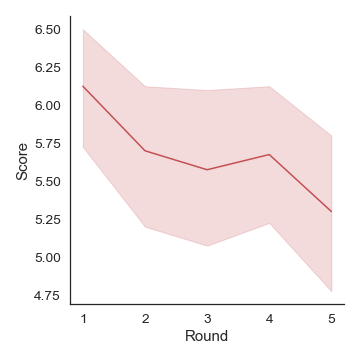}
 \end{multicols}
 \end{figure} 
\FloatBarrier
Neither regression coefficient is positive and significant. Thus, there is no evidence of learning effects.

\newpage
\subsection{Machine-learning analysis to detect heterogeneous treatment effects}

Yet, we might have heterogeneous treatment effects. To see if there is any evidence for this, we then construct a non-parametric Gaussian mixture model, and apply it to the overconfident subjects' results from the first and fifth round.\footnote{We built and implemented the algorithm in Python using the \textit{Scikit Learn} package - see $https://scikit-learn.org/$}

 A Gaussian mixture model is an algorithm that divides a set of data into a number of clusters. The modeller takes a set of data, and specifies a number of clusters to divide the data into. The algorithm then finds the clusters that best describes the information in the data, using an expectation-maximisation algorithm.\footnote{More technically, the modeller specifies the number of Gaussian distributions they think the data is drawn from. The algorithm then finds the number of distributions (clusters) that best describe the data. As we say above, we use a non-parametric version of the model. Therefore we fit models with a range of different numbers of distributions, and then select the model that captures the most information about the data using an information criterion.} To do this, you need to know the numbers of clusters to divide the data into, which we did not. Hence our algorithm takes a range of possible number of clusters (one to fifteen), clusters the data, and gives each assignment a score based on how well it describes the data. The algorithm then selects the model with the best score.\footnote{More technically, we calculate the \textit{`Bayesian Information'} score of each possible model, and minimise this. Our results do not depend on the information criterion we use - see below} 
 
 \begin{figure}[!h]

\centering

   \caption{Example Gaussian Mixture Model Output}
 \includegraphics[width=12cm]{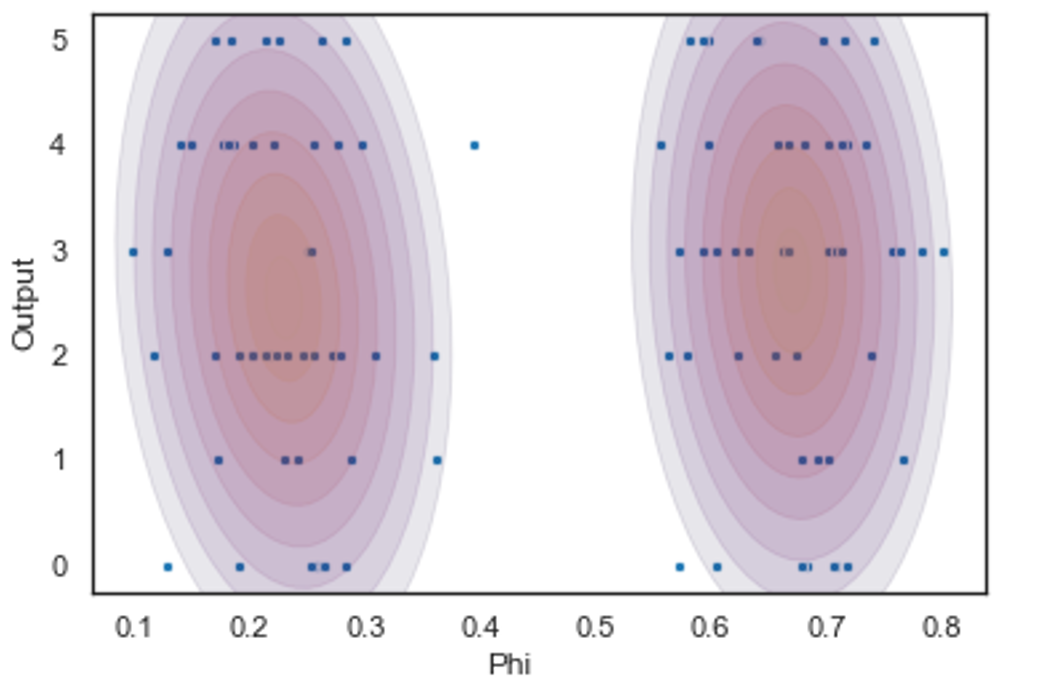}
 
 \smallskip
 
 \footnotesize
 Example of output we would observe if all subjects could be described using \cite{Heidhues2018}'s theory. Generated using synthetic data with $n=100$. Means and marks for each distribution are generated to make them two distributions as clear and distinct as possible.
 \end{figure}
 
  We have a set of points representing each subjects' scores and beliefs about $\phi$ in the first and fifth rounds.  If updating does cause an overconfident subject's beliefs to fall over updating rounds, the point representing their score should move downwards in the $\phi$ dimension between the rounds. Thus, if updating causes all overconfident subjects' beliefs about $\phi$ to fall, the pairs of results in the first and final round should appear to be drawn from two different distributions – one for subjects in the first round with a higher belief about $\phi$, and another for subjects from the final round with a lower belief about $\phi$ (e.g see Figure $5$). Now consider the case of heterogeneous treatment effects. If updating does cause some subjects' beliefs about $\phi$ to fall, but not the others, the algorithm should identify those subjects points as being drawn from two distributions, one higher in $\phi$ space and the other lower in $\phi$ space, in amongst noise. Using the Gaussian mixture model is a better way of identifying heterogeneous causal effects than simply looking at the distribution of treatment effects. Using the algorithm, we detect effects by finding the data generating process that best accounts for the information in the sample. But, if all of the individual treatment effects were drawn randomly, some may be large enough to appear significant by chance, not due to the underlying data generating process. Thus, if we just look at the individual effects, it may appear as if a proportion of our subjects followed Heidhues et al.'s model, even if they did not.  

 Yet, as shown in Figure $6$, the model that best fits our data does not look like this - individuals appear to be drawn from distributions centered on output, not beliefs about $\phi$.\footnote{Note that this is not just because there can be a greater ranges of output than $\phi$ values - see the Appendix.}\\

\begin{figure}[!h]
\centering
   \caption{Clusters in Data from Overconfident Subjects in Rounds $1$ and $5$}
 \includegraphics[width=12cm]{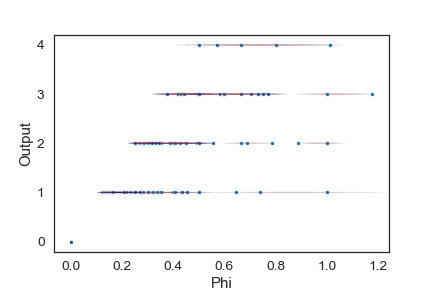}
 \end{figure}
Now,  we carry out some robustness checks.
The Gaussian mixture model might have clustered subjects as it has in Figure $6$ because subjects can get a greater range of marks than beliefs about $\phi$. Hence the Euclidean distance between subjects who have the same mark but different beliefs about $\phi$ will tend to be higher than between subjects who have the same beliefs about $\phi$ but different marks. So the algorithm might be more likely to cluster subjects by mark than by $\phi$.\\

To ensure that this is not why our algorithm is clustering subjects as in Figure $6$, we multiplied each subject's $\phi$ value by $10$. This means the Euclidean distance between subjects with different beliefs about $\phi$ but the same mark will tend to be higher than between subjects with different marks but the same belief about $\phi$. If the algorithm is just clustering subjects in Figure $6$ because of the different ranges of mark and $\phi$, the clusters will disappear.\\

We then ran the same algorithm to cluster this new data. The results are below:

 \begin{figure}[!h]
 \centering
   \caption{Clusters in Data from Overconfident Subjects in Rounds 1 and 5 - Transformed Data}
\label{figure:1}
\includegraphics[width=7.5cm]{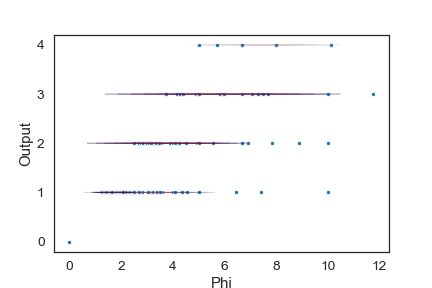}

\footnotesize Generated with the data from overconfident subjects in round 1 and 5, where all `Phi' values are multiplied by $10$

 \end{figure}

The clusters from Figure $6$ do not disappear. Hence there is no evidence that we find these clusters because marks and beliefs about $\phi$ are defined over different ranges.\\

The results could also just be because of the information criterion we use to select models - the Bayesian information criterion. To check for this, we re-ran the algorithm and used the alternative Akaike information criterion instead. The algorithm assigns all points to exactly the same clusters.\\
 
\newpage

\subsection{Survey Instructions and Design}

In this section, we go though our survey and show what subjects saw at each stage. This subsection includes the instructions subjects read. The next section presents the questions we asked subjects by round.\\

Instructions:
\FloatBarrier
 \begin{figure}[h]
 \centering
\label{figure:1}
\includegraphics[width=10cm]{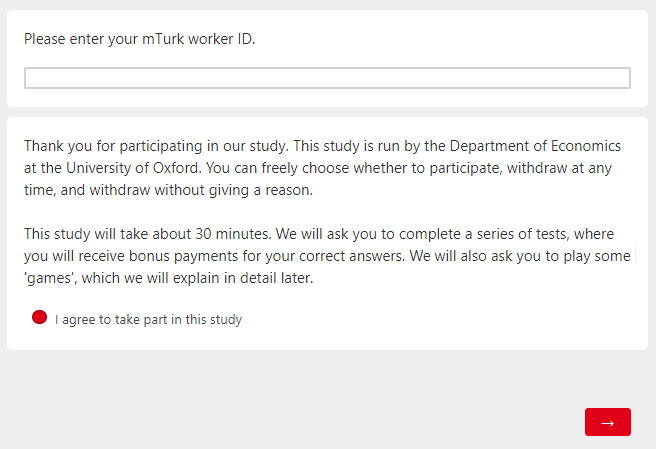}
\end{figure}

 \begin{figure}[h]
 \centering
\label{figure:1}
\includegraphics[width=10cm]{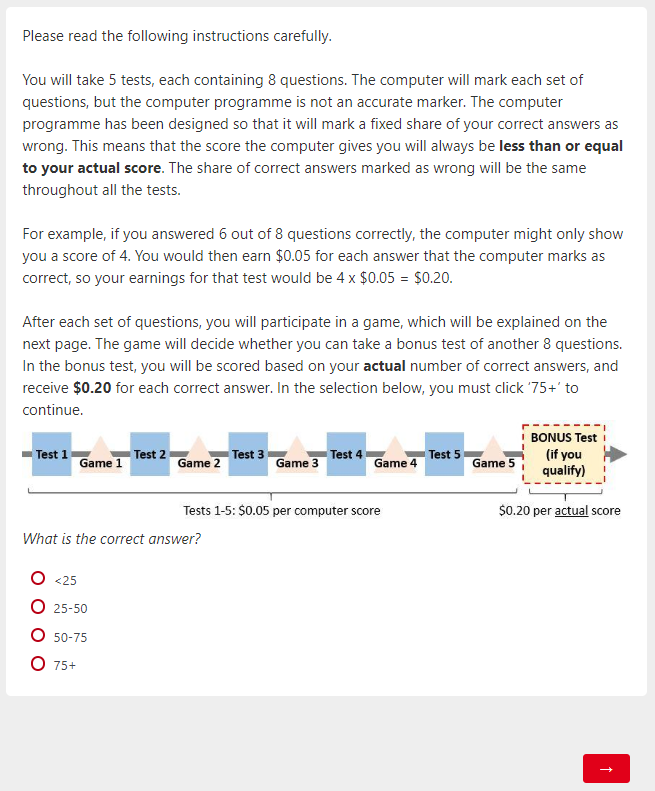}
\end{figure}

\FloatBarrier

If a subject failed the attention check, they were shown the following message and we ended the survey.

 \begin{figure}[h]
 \centering
\label{figure:1}
\includegraphics[width=10cm]{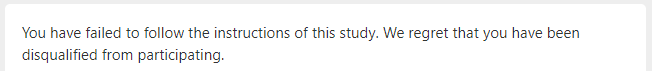}
\end{figure}

\FloatBarrier

\newpage

BDM instructions and test for understanding:\\

 \begin{figure}[h]
 \centering
\label{figure:1}
\includegraphics[width=10cm]{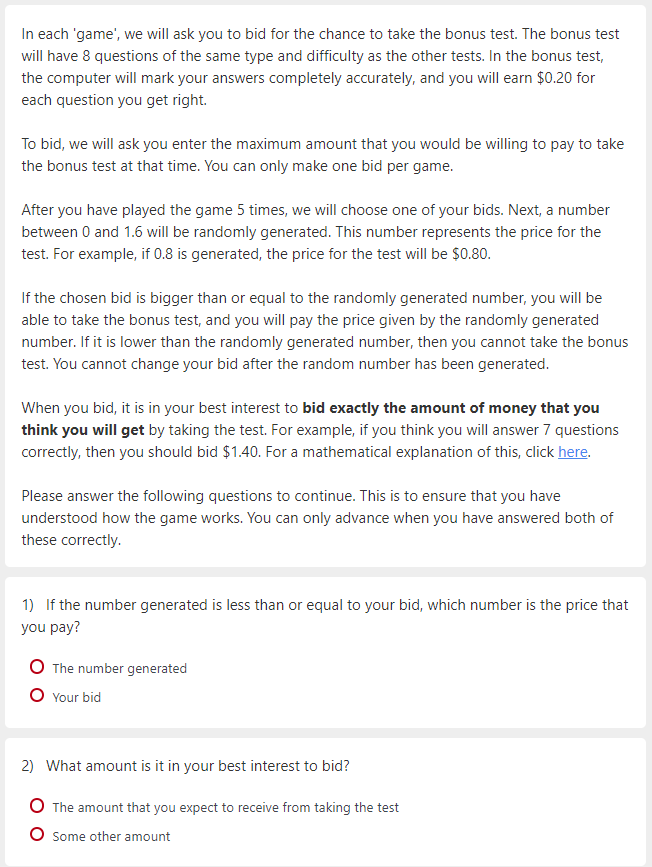}
\end{figure}

\FloatBarrier

\newpage

Subjects then took $5$ rounds of tests. These are the screens subjects saw during the first round:\\

\FloatBarrier

 \begin{figure}[h]
 \centering
\label{figure:1}
\includegraphics[width=10cm]{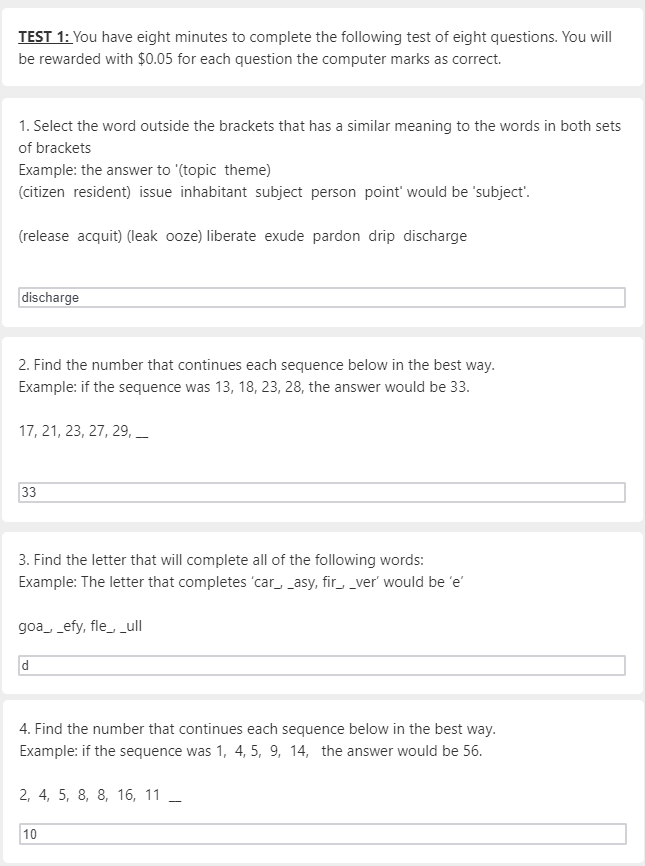}
\end{figure}

 \begin{figure}[h]
 \centering
\label{figure:1}
\includegraphics[width=10cm]{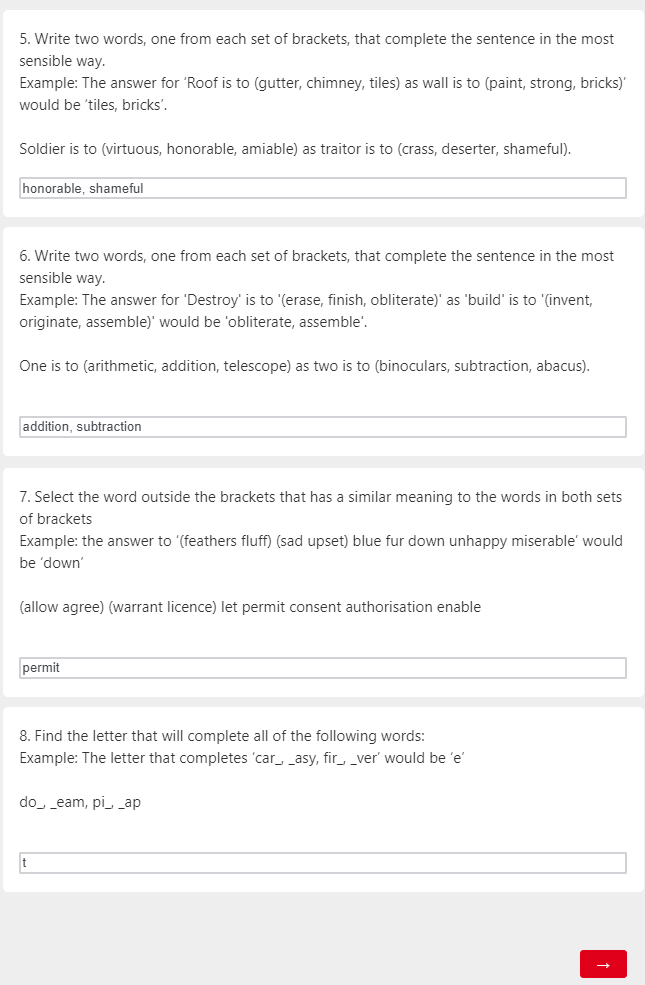}
\end{figure}

 \begin{figure}[h]
 \centering
\label{figure:1}
\includegraphics[width=10cm]{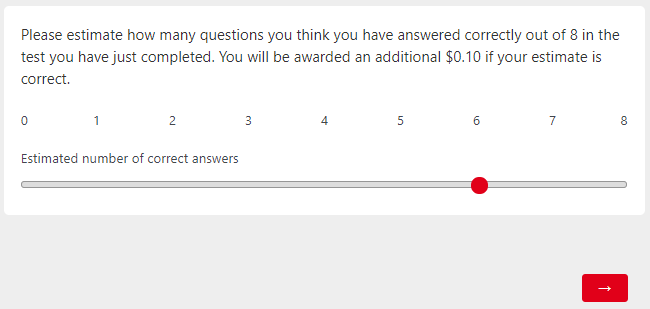}
\end{figure}

 \begin{figure}[h]
 \centering
\label{figure:1}
\includegraphics[width=10cm]{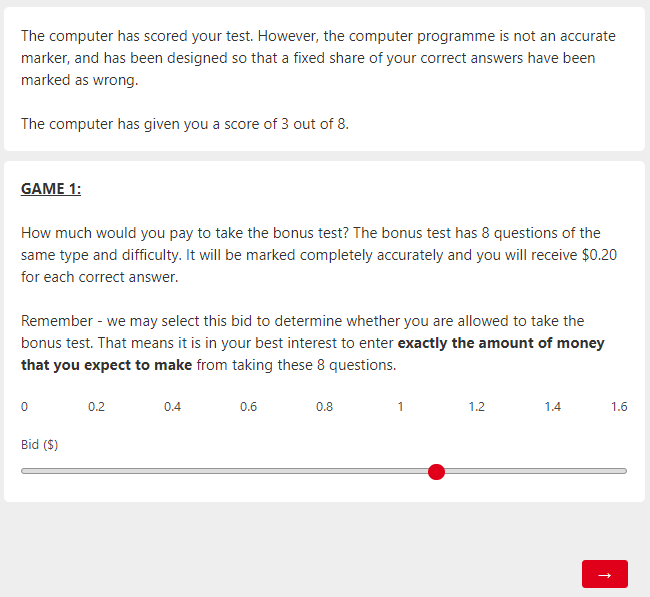}
\end{figure}

\FloatBarrier

After they had completed all of the rounds, a bid was then selected:\\

\FloatBarrier

 \begin{figure}[h]
 \centering
\label{figure:1}
\includegraphics[width=10cm]{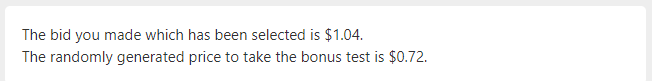}
\end{figure}

\FloatBarrier

If they did not win the extra test, subjects were shown this screen:\\

\FloatBarrier
 \begin{figure}[h]
 \centering
\label{figure:1}
\includegraphics[width=10cm]{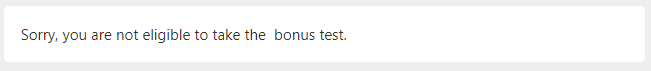}
\end{figure}
\FloatBarrier
If not, then they took the bonus test as below:\\
\FloatBarrier

 \begin{figure}[h]
 \centering
\label{figure:1}
\includegraphics[width=10cm]{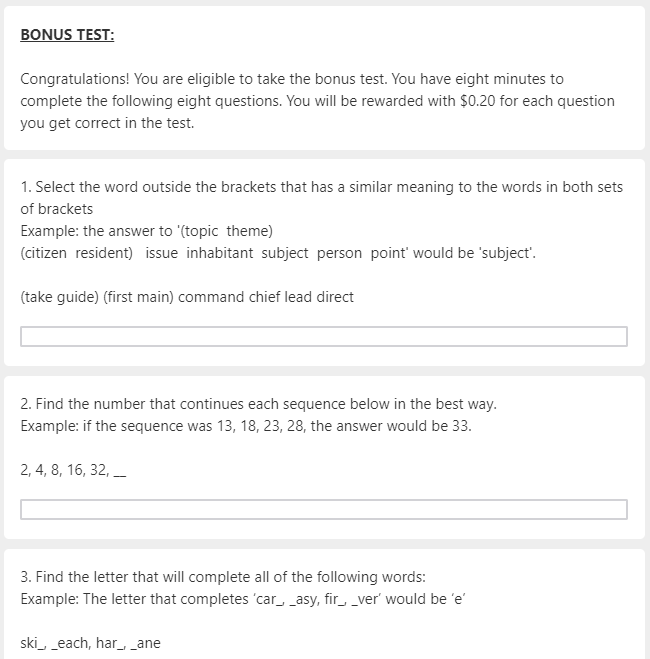}
\end{figure}

\FloatBarrier

\newpage

At the end of the experiment, subjects were first asked to complete some questions before the end.\\ 

\FloatBarrier

 \begin{figure}[h]
 \centering
\label{figure:1}
\includegraphics[width=10cm]{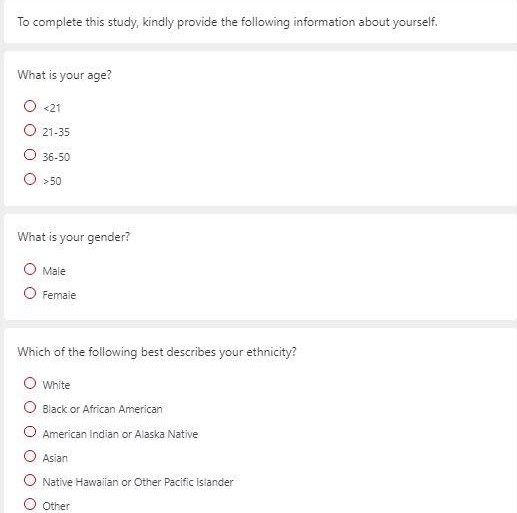}
\end{figure}
\FloatBarrier

\FloatBarrier

 \begin{figure}[h]
 \centering
\label{figure:1}
\includegraphics[width=10cm]{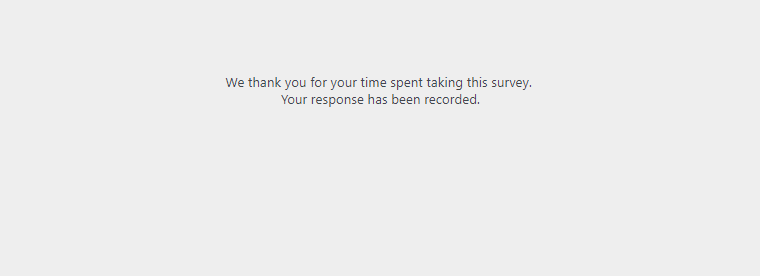}
\end{figure}
\FloatBarrier

\newpage
\subsubsection{Test Questions}

Questions used by round:\\

Round $1$\\

$1$. Select the word outside the brackets that has a similar meaning to the words in both sets of brackets\\

Example: the answer to '(topic  theme) (citizen  resident)  issue  inhabitant  subject  person  point' would be 'subject'.\\

(release  acquit) (leak  ooze) liberate  exude  pardon  drip  discharge\\

$2$. Find the number that continues each sequence below in the best way.

Example: if the sequence was $13, 18, 23, 28$, the answer would be $33$.\\

$17, 21, 23, 27, 29$,\\

3. Find the letter that will complete all of the following words:\\

Example: The letter that completes ‘car$\_$ , $\_$asy, fir$\_$, $\_$ver’ would be ‘e’\\

goa$\_$, $\_$efy, fle$\_$, $\_$ull\\

4. Find the number that continues each sequence below in the best way.\\

Example: if the sequence was $1,  4, 5,  9,  14$, the answer would be $56$.\\

$2,  4,  5,  8,  8,  16,  11$\\

5. Write two words, one from each set of brackets, that complete the sentence in the most sensible way.\\

Example: The answer for ‘Roof is to (gutter, chimney, tiles) as wall is to (paint, strong, bricks)’ would be ‘tiles, bricks’.\\

Soldier is to (virtuous, honorable, amiable) as traitor is to (crass, deserter, shameful).\\

6. Write two words, one from each set of brackets, that complete the sentence in the most sensible way.\\

Example: The answer for 'Destroy' is to '(erase, finish, obliterate)' as 'build' is to '(invent, originate, assemble)' would be 'obliterate, assemble'.\\ 

One is to (arithmetic, addition, telescope) as two is to (binoculars, subtraction, abacus).\\

7. Select the word outside the brackets that has a similar meaning to the words in both sets of brackets\\

Example: the answer to ‘(feathers fluff) (sad upset) blue fur down unhappy miserable’ would be ‘down’\\

(allow agree) (warrant licence) let permit consent authorisation enable\\

8. Find the letter that will complete all of the following words:\\

Example: The letter that completes ‘car$\_$, $\_$asy, fir$\_$, $\_$ver’ would be ‘e’\\

do$\_$, $\_$eam, pi$\_$, $\_$ap\\

Round 2\\

1. Select the word outside the brackets that has a similar meaning to the words in both sets of brackets.\\

Example: the answer to '(topic  theme) (citizen  resident) issue  inhabitant  subject  person  point' would be 'subject'.\\

(access  doorway) (delight  charm) entrance  portal  captivate  bewitch  gate.\\

 2. Find the number that continues each sequence below in the best way.\\

Example: if the sequence was $13, 18, 23, 28$, the answer would be $33$.\\

$14, 17,  19,  20,  20$\\

3. Find the letter that will complete all of the following words:\\

Example: The letter that completes ‘car$\_$, $\_$asy, fir$\_$, $\_$ver’ would be ‘e’.\\

loo$\_$, $\_$ole, for$\_$, $\_$elt\\ 

4. Write two words, one from each set of brackets, that complete the sentence in the most sensible way.\\

Example: The answer for ‘Roof is to (gutter, chimney, tiles) as wall is to (paint, strong, bricks)’ would be ‘tiles, bricks’.\\

Heart is to (circulation, blood, arteries) as lungs is to (organ, air, throat).\\

5. Select the word outside the brackets that has a similar meaning to the words in both sets of brackets.\\

Example: the answer to '(topic  theme) (citizen  resident)   issue  inhabitant  subject  person  point' would be 'subject'.\\

(error fault) (muddle confuse) wrong mistake puzzle baffle slip\\

6. Find the number that continues each sequence below in the best way.\\

Example: if the sequence was $13, 18, 23, 28$ the answer would be $33$.\\

$1, 4, 9, 16, 25$\\

7. Find the letter that will complete all of the following words:\\

Example: The letter that completes ‘car$\_$, $\_$asy, fir$\_$, $\_$ver’ would be ‘e’.\\

min$\_$, $\_$ar, se$\_$, $\_$ast\\

8. Write two words, one from each set of brackets, that complete the sentence in the most sensible way.\\

Example: The answer for ‘Roof is to (gutter, chimney, tiles) as wall is to (paint, strong, bricks)’ would be ‘tiles, bricks’.\\

Dull is to (dim slow dirty) as light is to (bright beacon coloured)\\

Round $3$\\

1. Select the word outside the brackets that has a similar meaning to the words in both sets of brackets\\

Example: the answer to '(topic  theme) (citizen  resident)   issue  inhabitant  subject  person  point' would be 'subject'.\\

(dry clear) (nice lovely) bright good great fine\\
 
2. Find the number that continues each sequence below in the best way.\\

Example: if the sequence was $13, 18, 23, 28$ the answer would be $33$.
$4, 10, 16, 22, 28$,\\

3. Find the letter that will complete all of the following words:\\

Example: The letter that completes ‘car$\_$, $\_$asy, fir$\_$, $\_$ver’ would be ‘e’\\

bra$\_$, $\_$est, war$\_$, $\_$ose\\

4. Write two words, one from each set of brackets, that complete the sentence in the most sensible way.\\

Example: The answer for ‘Roof is to (gutter, chimney, tiles) as wall is to (paint, strong, bricks)’ would be ‘tiles, bricks’.\\

Smell is to (sense, sneeze, nose) as hear is to (listen, ear, noise).\\
 
5. Select the word outside the brackets that has a similar meaning to the words in both sets of brackets.\\

Example: the answer to '(topic  theme) (citizen  resident)   issue  inhabitant  subject  person  point' would be 'subject'.\\

(near adjacent) (shut secure) lock adjoining close seal neighbouring\\

6. Find the number that continues each sequence below in the best way.\\

Example: if the sequence was $13, 18, 23, 28$ the answer would be $33$\\.

$6, 5, 5, 6, 8,$\\

7. Find the letter that will complete all of the following words:\\

Example: The letter that completes ‘car$\_$, $\_$asy, fir$\_$, $\_$ver’ would be ‘e’.\\

bar$\_$, $\_$ey, mar$\_$, $\_$it\\

8. Write two words, one from each set of brackets, that complete the sentence in the most sensible way.\\

Example: The answer for ‘Roof is to (gutter, chimney, tiles) as wall is to (paint, strong, bricks)’ would be ‘tiles, bricks’.\\

Car is to (wheel, engine, oil) as carriage is to (axle, horse, gilded)\\

Round $4$\\

1. Select the word outside the brackets that has a similar meaning to the words in both sets of brackets\\

Example: the answer to '(topic  theme) (citizen  resident)   issue  inhabitant  subject  person  point' would be 'subject'.\\

(turn revolve) (bun bread) snack spin circle roll.\\
 
2. Find the number that continues each sequence below in the best way.\\

Example: if the sequence was $13, 18, 23, 28$ the answer would be $33$.\\

$1, 10, 19, 28, 37$,\\

3. Find the letter that will complete all of the following words:\\

Example: The letter that completes ‘car$\_$, $\_$asy, fir$\_$, $\_$ver’ would be ‘e’.\\

he$\_$, $\_$ain, pe$\_$, $\_$ed\\

4. Write two words, one from each set of brackets, that complete the sentence in the most sensible way.\\

Example: The answer for ‘Roof is to (gutter, chimney, tiles) as wall is to (paint, strong, bricks)’ would be ‘tiles, bricks’.\\

Paint is to (easel artist brush) as stone is to (catapult wall sculptor).\\
 
5. Select the word outside the brackets that has a similar meaning to the words in both sets of brackets.\\

Example: the answer to '(topic  theme) (citizen  resident)   issue  inhabitant  subject  person  point' would be 'subject'.\\

(article item) (aim end) object motive goal thing gadget\\

6. Find the number that continues each sequence below in the best way.\\

Example: if the sequence was $13, 18, 23, 28$ the answer would be $33$.\\

$8, 9, 13, 7, 18, 5$,\\

7. Find the letter that will complete all of the following words:\\

Example: The letter that completes ‘car$\_$, $\_$asy, fir$\_$, $\_$ver’ would be ‘e’.\\

le$\_$, $\_$low, fla$\_$, $\_$un\\

8. Write two words, one from each set of brackets, that complete the sentence in the most sensible way.\\

Example: The answer for ‘Roof is to (gutter, chimney, tiles) as wall is to (paint, strong, bricks)’ would be ‘tiles, bricks’.\\

Doctor is to (hospital, medicine, nurse) as solicitor is to (client, contract, law)\\

Round 5\\

1. Select the word outside the brackets that has a similar meaning to the words in both sets of brackets\\

Example: the answer to '(topic  theme) (citizen  resident)   issue  inhabitant  subject  person  point' would be 'subject'.\\

(pink blush) (soared ascended) red rose bloom floated\\
 
2. Find the number that continues each sequence below in the best way.\\

Example: if the sequence was $13, 18, 23, 28$ the answer would be $33$.\\

$1, 2, 3, 5, 8$,

3. Find the letter that will complete all of the following words:\\

Example: The letter that completes ‘car$\_$, $\_$asy, fir$\_$, $\_$ver’ would be ‘e’.\\

ra$\_$, $\_$et, bu$\_$, $\_$east\\

4. Write two words, one from each set of brackets, that complete the sentence in the most sensible way.\\

Example: The answer for ‘Roof is to (gutter, chimney, tiles) as wall is to (paint, strong, bricks)’ would be ‘tiles, bricks’.\\

Teacher is to (desk, classroom, school) as professor is to (university, lecture, study).\\
 
5. Select the word outside the brackets that has a similar meaning to the words in both sets of brackets.\\

Example: the answer to '(topic  theme) (citizen  resident)   issue  inhabitant  subject  person  point' would be 'subject'.\\

(award medal) (value cherish) bonus reward treasure love prize\\

6. Find the number that continues each sequence below in the best way.\\

Example: if the sequence was $13, 18, 23, 28$ the answer would be $33$.\\

$37, 35, 31, 25, 17$,\\

7. Find the letter that will complete all of the following words:\\

Example: The letter that completes ‘car$\_$, $\_$asy, fir$\_$, $\_$ver’ would be ‘e’.\\

bo$\_$, $\_$ay, se$\_$, $\_$it\\

8. Write two words, one from each set of brackets, that complete the sentence in the most sensible way.\\

Example: The answer for ‘Roof is to (gutter, chimney, tiles) as wall is to (paint, strong, bricks)’ would be ‘tiles, bricks’.\\

Apple is to (crunchy, orchard, core) as grape is to (sweet, wine, vineyard)\\

Bonus Round\\

1. Select the word outside the brackets that has a similar meaning to the words in both sets of brackets\\

Example: the answer to '(topic  theme) (citizen  resident)   issue  inhabitant  subject  person  point' would be 'subject'.\\ 

(take guide) (first main) command chief lead direct\\

2. Find the number that continues each sequence below in the best way.\\

Example: if the sequence was $13, 18, 23, 28$ the answer would be $33$.\\

$2, 4, 8, 16, 32$,\\

3. Find the letter that will complete all of the following words:\\

Example: The letter that completes ‘car$\_$, $\_$asy, fir$\_$, $\_$ver’ would be ‘e’.\\

ski$\_$, $\_$each, har$\_$, $\_$ane\\

4. Find the number that continues each sequence below in the best way.\\

Example: if the sequence was $13, 18, 23, 28$ the answer would be $33$.\\

$16, 2, 14, 6, 12, 10$,\\

5. Write two words, one from each set of brackets, that complete the sentence in the most sensible way.\\

Example: The answer for ‘Roof is to (gutter, chimney, tiles) as wall is to (paint, strong, bricks)’ would be ‘tiles, bricks’.\\

Tractor is to (trailer plough farmer) as tank is to (fish soldier cannon).\\

6. Write two words, one from each set of brackets, that complete the sentence in the most sensible way.\\

Example: The answer for 'Destroy' is to '(erase, finish, obliterate)' as 'build' is to '(invent, originate, assemble)' would be 'obliterate, assemble'.\\ 

Bread is to (wheat roll knead) as butter is to (churn dish cheese). \\

7. Select the word outside the brackets that has a similar meaning to the words in both sets of brackets\\

Example: the answer to ‘(feathers fluff) (sad upset) blue fur down unhappy miserable’ would be ‘down’\\

(firm solid) (difficult awkward) rigid complex stiff troublesome hard\\

8. Find the letter that will complete all of the following words:\\

Example: The letter that completes ‘car$\_$, $\_$asy, fir$\_$, $\_$ver’ would be ‘e’.\\

mil$\_$, $\_$ong, pee$\_$, $\_$oot

\end{document}